\documentclass[10pt,a4paper]{article}
\usepackage{amsmath,amssymb,amsthm}
\usepackage{mathptmx}
\usepackage[scaled=.92]{helvet}
\usepackage[margin=2.5cm]{geometry}
\usepackage{graphicx}
\usepackage{tikz}
\usetikzlibrary{arrows.meta,calc,decorations.pathreplacing}
\tikzset{
  ext/.style={circle,draw,fill=white,inner sep=0pt,minimum size=5pt,line width=.6pt},
  int/.style={circle,draw,fill=black,inner sep=0pt,minimum size=5pt},
  Wedge/.style={line width=.9pt},
  Wbar/.style={line width=.9pt,dash pattern=on 3pt off 2pt},
  lab/.style={font=\small},
}
\usepackage[numbers,sort&compress]{natbib}
\usepackage{titlesec}
\usepackage{caption}
\usepackage[colorlinks=true,linkcolor=black,citecolor=black,urlcolor=blue]{hyperref}

\titleformat{\section}{\centering\normalfont\bfseries}{\thesection.}{.5em}{\MakeUppercase}
\titleformat{\subsection}{\centering\normalfont\bfseries}{\thesubsection.}{.5em}{}
\titleformat{\paragraph}[runin]{\normalfont\bfseries}{}{0pt}{}[.]
\newtheorem{theorem}{Theorem}
\newtheorem{proposition}{Proposition}
\newtheorem{lemma}{Lemma}
\theoremstyle{remark}
\newtheorem{remark}{Remark}

\newcommand{\Qp}{\mathbb Q_p}
\newcommand{\Zp}{\mathbb Z_p}
\newcommand{\Rc}{\check{\mathcal R}}
\newcommand{\GL}{GL_2(\Qp)}

\begin{document}

\begin{center}
{\Large\bfseries Yang-Baxter Operator over the {\itshape p}-Adic Numbers}

\bigskip
{\large Ilmar Gahramanov$^{a,b}$ and Firdevs Karaku\c{s}$^{a,c,*}$}

\medskip
{\small\itshape
$^{a}$Department of Physics, Bo\u{g}azi\c{c}i University, Istanbul, 34342 T\"urkiye\\
$^{b}$Center for Mathematics and its Applications, Khazar University, Baku, AZ1096 Azerbaijan\\
$^{c}$Department of Mathematics, Bo\u{g}azi\c{c}i University, Istanbul, 34342 T\"urkiye}

\medskip
{\small
e-mail: \href{mailto:ilmar.gahramanov@bogazici.edu.tr}{ilmar.gahramanov@bogazici.edu.tr}, \href{mailto:firdevs.karakus@std.bogazici.edu.tr}{firdevs.karakus@std.bogazici.edu.tr}\\
$^*$Corresponding author}
\end{center}

\begin{abstract}
\noindent We study the star-triangle relation for power-law Boltzmann weights over $\Qp$. Normalizing the convolution weights by their Fourier multipliers removes the product of local gamma functions from the relation. The resulting operator identity realizes the Coxeter relations for permutations of the parameters. Our main result is the factorized Yang-Baxter operator obtained in this way. For imaginary spectral and spin parameters, this operator is unitary on $L^2(\Qp^2)$, satisfies the Yang-Baxter equation as an operator identity and intertwines the Hilbert tensor products of two unitary principal series representations of $\GL$ in opposite orders. At zero spectral parameter it is the identity for equal spins. The operator identity also gives a Bailey lemma and a transformation formula for $p$-adic four-point integrals. To apply Wu's edge-interaction construction of knot invariants, we average individual pair weights over the residue discs of the projective line. Invariance under $PGL_2(\Zp)$ gives a $(p+1)$-state Potts model. At exponents determined by the two finite inversion relations, the meromorphically continued weights yield the Jones polynomial at $t+t^{-1}=p-1$ and involve the unramified quadratic character of $\Qp^\times$. We also compute the off-diagonal leading term at the crossing point and discuss the regularization needed to relate it to the Temperley-Lieb algebra at zero loop parameter.

\medskip
\noindent\textbf{Keywords:} $p$-adic numbers, star-triangle relation, Yang-Baxter equation, principal series representations, Potts model, Jones polynomial

\medskip
\noindent\textbf{Mathematics Subject Classification:} 82B23, 16T25, 11S80, 22E50, 57K14
\end{abstract}

\section{Introduction}\label{sec:intro}

In recent years, there has been significant progress in the study of integral identities for special functions and their applications to integrable lattice models. In the so-called gauge/YBE correspondence, such identities arise from the equality of partition functions of supersymmetric dual theories and can be written as star-triangle relations for Boltzmann weights. This correspondence was developed by Yamazaki~\cite{Yamazaki:2012cp}, and it has led to the construction of many new integrable lattice spin models, see the reviews~\cite{Yamazaki:2018xbx,Mullahasanoglu:2025ont} and references therein. Closely related integrability conditions, such as the star-square, decoration and flipping relations, were obtained in the same framework~\cite{Mullahasanoglu:2023nes,Catak:2024ygo,Catak:2025hoz}.

In~\cite{Gahramanov:2015cva}, a basic hypergeometric sum-integral identity was related to the star-triangle relation and to the vertex and interaction-round-a-face forms of the Yang-Baxter equation. Solutions in terms of Euler's gamma function, which are close analogues of the power-law weights considered here, were obtained as limits of such identities in~\cite{Eren:2019ibl,Mullahasanoglu:2024stv}. The star-triangle relation is also related to the pentagon identity, i.e.\ to the $2$-$3$ Pachner move~\cite{Yamazaki:2012cp,Bozkurt:2020gyy}, and Bailey-pair constructions for these identities were developed in~\cite{Gahramanov:2022jxz}. A useful tool in this approach is the operator form of the star-triangle relation, which allows one to construct Yang-Baxter operators through the Coxeter relations of the permutation group of the parameters~\cite{Gahramanov:2015cva,Gahramanov:2022jxz}. Such factorized operators were constructed by Derkachov and Manashov for noncompact $SL(N,\mathbb C)$ spin chains~\cite{Derkachov:2006fw,Derkachov:2008aq}, and by Chicherin, Derkachov and Isaev for the conformal group~\cite{Chicherin:2012yn}. In these works the star-triangle relation plays a prominent role, and the Yang-Baxter operator has a natural interpretation as an intertwiner of tensor products of principal series representations. For the real principal series, Neretin~\cite{Neretin:2024} gave a rigorous operator proof using a deformation of the star-triangle identity.

Over the last decades, $p$-adic numbers have found many applications in mathematical physics. $p$-Adic mathematical physics was initiated by Volovich in 1987~\cite{Volovich:1987nq,Volovich:1987zc}, who proposed considering string theory over the field of $p$-adic numbers. In this approach the string amplitudes are given by $p$-adic integrals of products of power functions, which can be evaluated in terms of the local gamma function~\cite{Volovich:1987nq,Volovich:1988rs}. The $p$-adic Veneziano amplitude was also studied by Freund and Olson~\cite{Freund:1987kt}, and its product over all primes, i.e.\ the adelic formula, was considered by Freund and Witten~\cite{Freund:1987ck}. Zabrodin gave a geometric interpretation of $p$-adic strings in terms of the Bruhat-Tits tree~\cite{Zabrodin:1988ep}. Later the subject was developed further by Vladimirov and Volovich, in particular in $p$-adic quantum mechanics~\cite{Vladimirov:1988bd}. Non-Archimedean analogues of conformal field theory were introduced by Melzer~\cite{Melzer:1988he} at the end of the 1980s, and they appeared again in the context of the $p$-adic AdS/CFT correspondence~\cite{Gubser:2016guj}. The interested reader is referred to the monograph~\cite{Vladimirov:1994wi} and the reviews~\cite{Brekke:1993gf,Dragovich:2009hd,Dragovich:2017kge} for more details.

In this paper we consider a $p$-adic version of the operator construction discussed above. The point of departure is the power-law integral identity over $\Qp$ discussed by Gubser, Jepsen, Parikh and Trundy~\cite{Gubser:2017vgc}. As noted there, this identity generalizes the integral which gives the Veneziano amplitude, and its spin-independent factor is nothing but the $p$-adic Veneziano amplitude. We write this identity in terms of Boltzmann weights and apply the operator method of~\cite{Gahramanov:2015cva}. In our case the spin variables take values in $\Qp$, and the local gamma function plays the role of the special function which appears in the normalization of the weights.

The main novelty of this work is the construction of a unitary Yang-Baxter operator from the $p$-adic star-triangle relation. Its integral kernel is
\begin{equation}\label{eq:intro-R}
 \Rc(x_1,x_2;z_1,z_2)=
 \frac{|x_1-x_2|_p^{-\delta-g_1-g_2}\,|x_1-z_1|_p^{\delta+g_1-g_2-1}\,|x_2-z_2|_p^{\delta-g_1+g_2-1}\,|z_1-z_2|_p^{-\delta+g_1+g_2}}
 {\Gamma_p(\delta+g_1-g_2)\,\Gamma_p(\delta-g_1+g_2)},
\end{equation}
where $\delta$ stands for the spectral parameter, $g_1,g_2$ are the spin parameters and $\Gamma_p$ is the local gamma function~\eqref{eq:gamma}. The kernel is a product of four power-law Boltzmann weights. The operator $\Rc_{12}(\delta;g_1,g_2)$ with this kernel factorizes into two multiplication operators and two convolution operators, see~\eqref{eq:R}. For imaginary parameters the convolution factors are defined as Fourier multipliers, and all four factors are unitary. Our main result is the following theorem.

\begin{theorem}\label{thm:main}
Let $\delta,g_1,g_2\in i\mathbb R$ and $\rho_j=\tfrac12+g_j$. Then $\Rc_{12}(\delta;g_1,g_2)$ is a unitary operator on $L^2(\Qp^2)$ with the following properties.
\begin{itemize}
 \item[(i)] Yang-Baxter equation. For $g_3\in i\mathbb R$ and $u_1,u_2,u_3$ with $u_j-u_k\in i\mathbb R$,
 \begin{equation}\label{eq:intro-ybe}
 \begin{split}
 \Rc_{23}(u_1-u_2;g_1,g_2)\,&\Rc_{12}(u_1-u_3;g_1,g_3)\,\Rc_{23}(u_2-u_3;g_2,g_3)\\
 &=\Rc_{12}(u_2-u_3;g_2,g_3)\,\Rc_{23}(u_1-u_3;g_1,g_3)\,\Rc_{12}(u_1-u_2;g_1,g_2)
 \end{split}
 \end{equation}
 as an identity between unitary operators on $L^2(\Qp^3)$.
 \item[(ii)] Intertwining property. For all $h\in\GL$,
 \begin{equation}\label{eq:intro-int}
 \Rc_{12}(\delta;g_1,g_2)\,\bigl(T_{\rho_1}\otimes T_{\rho_2}\bigr)(h)
 =\bigl(T_{\rho_2}\otimes T_{\rho_1}\bigr)(h)\,\Rc_{12}(\delta;g_1,g_2),
 \end{equation}
 where $T_\rho$ is the unitary principal series representation of $\GL$ induced from the character $|\cdot|_p^{\rho}$. In other words, $\Rc_{12}$ intertwines the Hilbert tensor product $\mathcal H_{\rho_1}\widehat\otimes\mathcal H_{\rho_2}$ with the same product in the opposite order.
 \item[(iii)] Unitarity. $\Rc(\delta;g_1,g_2)^\dagger=\Rc(\delta;g_1,g_2)^{-1}=\Rc(-\delta;g_2,g_1)$.
 \item[(iv)] Initial condition. $\Rc(0;g,g)=1$.
\end{itemize}
\end{theorem}

The theorem is proved in Section~\ref{sec:yb}. The Yang-Baxter equation follows from the Coxeter relations of the permutation group of the parameters, and the only input is the operator form of the star-triangle relation, which we prove for imaginary parameters in Lemma~\ref{lem:opstr}. The intertwining property is Proposition~\ref{prop:intertwiner}, and the unitarity and the initial condition are proved in Section~\ref{sec:properties}. Theorem~\ref{thm:main} is a $p$-adic analogue of the factorized $SL(2,\mathbb C)$ R-operator of Derkachov and Manashov~\cite{Derkachov:2006fw}. Note, however, that in the $p$-adic case each site carries a single spin parameter $g_j$, and the spectral parameter $\delta$ is not a representation label.

As in~\cite{Gahramanov:2015cva,Gahramanov:2022jxz}, the same operator form of the star-triangle relation also leads to a Bailey lemma. Starting from an explicit seed, we obtain a transformation formula for $p$-adic four-point integrals.

Since the star-triangle relation is the local form of the third Reidemeister move, it is natural to ask whether the same Boltzmann weights lead to a knot invariant. The statistical-mechanical construction of knot invariants goes back to Jones~\cite{Jones:1989ed} and Kauffman~\cite{Kauffman:1987}. Akutsu, Deguchi and Wadati developed constructions of link polynomials from solvable vertex and IRF models~\cite{Akutsu:1987qs,Akutsu:1987qz,Akutsu:1987bc}, see~\cite{Wadati:1989ud} for a review. The question of whether $p$-adic knot invariants exist was raised by Morozov~\cite{Morozov:2015nra}, who proposed defining $p$-adic HOMFLY-PT polynomials for torus knots by using the Hall-Littlewood version of the Rosso-Jones formula. In this paper we follow a different route and use the edge-interaction formulation of Wu~\cite{Wu:1992zzb}. Besides the star-triangle relation, this construction requires two inversion relations and a normalization associated with the first Reidemeister move. The continuous power-law weights cannot satisfy both inversion relations (see Appendix~\ref{app:continuous}). After averaging the pair weights over the first residue level of the projective line $\mathbb P^1(\Qp)$, however, one obtains a finite Potts model, and Wu's conditions can be satisfied at
\begin{equation}\label{eq:intro-result}
 t_p+t_p^{-1}=p-1.
\end{equation}
The relation $t+t^{-1}=q-2$ between the number of states and the Jones variable is well known for the Potts model~\cite{Jones:1989ed,Wu:1992zzb}. Therefore the $p$-adic input is the number $q=p+1$ of residue discs together with the exponents of the weights. Interestingly, the averaged weights with exponents $a$ and $2-a$, i.e.\ the averaged analogue of a convolution-inverse pair of power functions, satisfy the off-diagonal part of the convolution relation identically, and the pointwise relation then fixes $a$. The resulting exponents contain the unramified quadratic character of $\Qp^\times$. Since the averaging is performed on individual pair weights, the relations of the finite model have to be checked directly.

The paper is organized as follows. In Section~\ref{sec:str}, we briefly recall the $p$-adic star-triangle relation and fix the normalization of the Boltzmann weights. In Section~\ref{sec:yb}, we construct the Yang-Baxter operator and prove Theorem~\ref{thm:main}. In Section~\ref{sec:bailey}, we derive a Bailey lemma from the operator form of the star-triangle relation and use it to obtain a transformation formula for four-point integrals. In Section~\ref{sec:jones}, we apply Wu's construction to a finite model obtained from the $p$-adic weights and show that it gives the Jones polynomial at~\eqref{eq:intro-result}. We conclude with a few remarks and open problems in Section~\ref{sec:concl}. In Appendix~\ref{app:unitary} we prove the operator form of the star-triangle relation in the unitary case. In Appendix~\ref{app:proof} we give the proof of the intertwining relations, and in Appendix~\ref{app:wu} we review Wu's construction. In Appendix~\ref{app:continuous} we show that the continuous weights do not satisfy Wu's relations. In Appendix~\ref{app:crossing} we study the crossing-point limit of the Yang-Baxter operator and its possible relation with the Temperley-Lieb algebra at zero loop parameter. We have also checked the star-triangle relation, the intertwining relations and the four-point transformation formula numerically for several small primes. These checks are not presented here.

\section{Star-triangle relation over \texorpdfstring{$\Qp$}{Qp}}\label{sec:str}

Let us briefly recall some basic definitions and refer the reader to~\cite{Gelfand:1969,Tate:1967,Vladimirov:1994wi} for more details. Let $p$ be a prime. We normalize the Haar measure on $\Qp$ by $\int_{\Zp}dx=1$ and choose the additive character $\psi(x)=e^{2\pi i\{x\}_p}$, where $\{x\}_p$ stands for the fractional part of $x$. The local gamma function is defined as
\begin{equation}\label{eq:gamma}
 \Gamma_p(a)=\frac{1-p^{a-1}}{1-p^{-a}},
\end{equation}
and satisfies the reflection formula
\begin{equation}\label{eq:reflection}
 \Gamma_p(a)\Gamma_p(1-a)=1.
\end{equation}
The normalization of $\Gamma_p$ follows from the Fourier transform of the power function~\cite{Gelfand:1969,Tate:1967,Vladimirov:1994wi},
\begin{equation}\label{eq:fourier}
 \int_{\Qp}|x|_p^{a-1}\psi(kx)\,dx=\Gamma_p(a)\,|k|_p^{-a}.
\end{equation}
The scalar power-law identities below extend meromorphically in their exponents. We specify the domains of the operator identities in Section~\ref{sec:yb}.

The integral identity which we use has the following form~\cite{Gubser:2017vgc},
\begin{equation}\label{eq:rawstr}
 \int_{\Qp}\prod_{j=1}^3|x_j-x_0|_p^{a_j-1}\,dx_0
 =\prod_{j=1}^3\Gamma_p(a_j)\;
 |x_{23}|_p^{-a_1}|x_{13}|_p^{-a_2}|x_{12}|_p^{-a_3},
\end{equation}
where $x_{ij}=x_i-x_j$, the external spins are distinct, and the exponents satisfy the balancing condition
\begin{equation}
 a_1+a_2+a_3=1.
\end{equation}
The integral converges for $\operatorname{Re}a_j>0$.

Let us now write this identity as a star-triangle relation. For this purpose we introduce the Boltzmann weight
\begin{equation}\label{eq:W}
 W_a(x,y)=|x-y|_p^{-a}.
\end{equation}
Then the identity~\eqref{eq:rawstr} takes the standard form of the star-triangle relation (see e.g.~\cite{Bazhanov:2016ajm} and references therein),
\begin{equation}\label{eq:str-R}
\begin{split}
 \int_{\Qp}W_{1-a_1}(x_1,x_0)\,&W_{1-a_2}(x_2,x_0)\,W_{1-a_3}(x_3,x_0)\,dx_0\\
 &=R(a_1,a_2,a_3)\,W_{a_1}(x_2,x_3)\,W_{a_2}(x_1,x_3)\,W_{a_3}(x_1,x_2),
\end{split}
\end{equation}
where $a_1+a_2+a_3=1$ and the factor $R$ has the following form
\begin{equation}\label{eq:R-factor}
 R(a_1,a_2,a_3)=\Gamma_p(a_1)\,\Gamma_p(a_2)\,\Gamma_p(a_3).
\end{equation}
Note that $R$ does not depend on the spins. Actually, it is nothing but the $p$-adic Veneziano amplitude~\cite{Volovich:1987nq,Freund:1987kt,Brekke:1993gf}, see also the discussion in~\cite{Gubser:2017vgc}. Indeed, setting $x_1=0$, $x_2=1$ in~\eqref{eq:rawstr}, multiplying both sides by $|x_3|_p^{1-a_3}$ and sending $x_3\to\infty$, one finds the well-known formula
\begin{equation}\label{eq:veneziano}
 \int_{\Qp}|x|_p^{a_1-1}\,|1-x|_p^{a_2-1}\,dx=\Gamma_p(a_1)\,\Gamma_p(a_2)\,\Gamma_p(a_3),\qquad a_1+a_2+a_3=1.
\end{equation}
In this sense the star-triangle relation~\eqref{eq:str-R} may be regarded as a spin-dependent extension of the $p$-adic beta integral. The parameter $1$ plays the role of the crossing parameter, i.e.\ the weights on the star side are obtained from the weights on the triangle side by the replacement $a_j\to1-a_j$.\footnote{The integration measure is the Haar measure, so the spin-dependent density which usually appears in the integrand of the star-triangle relation is trivial here.}

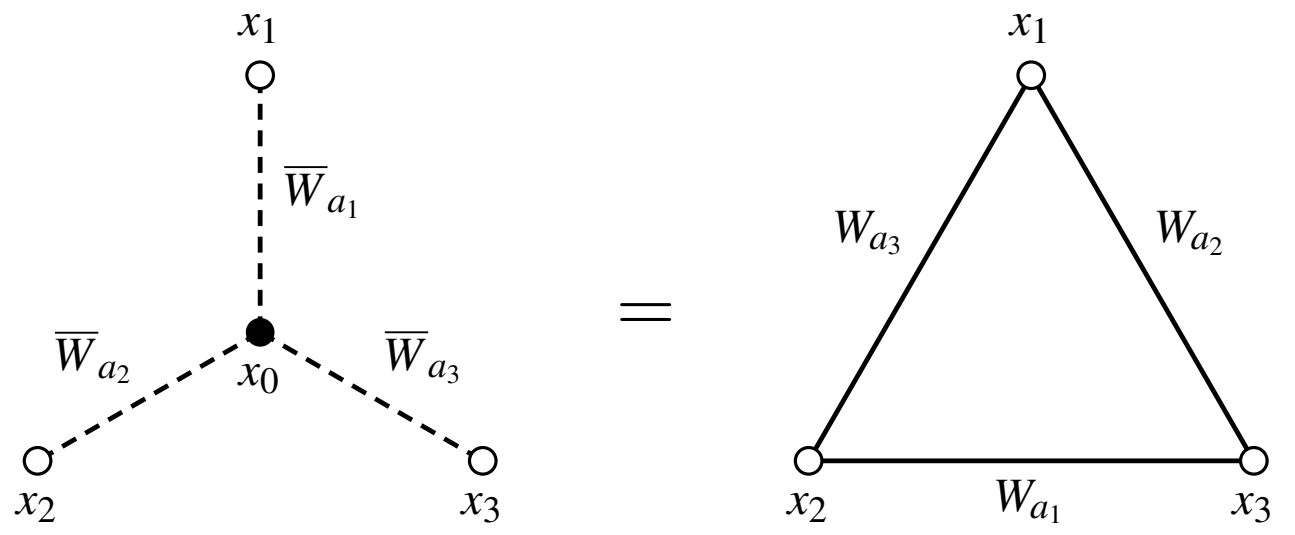
\begin{figure}[htbp]
\centering
{\sffamily
\begin{tikzpicture}
\begin{scope}
\coordinate (c) at (0,0);
\coordinate (v1) at (90:1.7); \coordinate (v2) at (210:1.7); \coordinate (v3) at (330:1.7);
\draw[Wbar] (c)--(v1) node[lab,pos=.55,right=1pt]{$\overline W_{a_1}$};
\draw[Wbar] (c)--(v2) node[lab,pos=.5,above left=-1pt]{$\overline W_{a_2}$};
\draw[Wbar] (c)--(v3) node[lab,pos=.5,above right=-1pt]{$\overline W_{a_3}$};
\node[int,label={[lab]below:$x_0$}] at (c){};
\node[ext,label={[lab]above:$x_1$}] at (v1){};
\node[ext,label={[lab]below:$x_2$}] at (v2){};
\node[ext,label={[lab]below:$x_3$}] at (v3){};
\end{scope}
\node at (2.55,0.1){\Large $=$};
\begin{scope}[xshift=5.1cm]
\coordinate (v1) at (90:1.7); \coordinate (v2) at (210:1.7); \coordinate (v3) at (330:1.7);
\draw[Wedge] (v2)--(v3) node[lab,midway,below]{$W_{a_1}$};
\draw[Wedge] (v1)--(v3) node[lab,midway,above right=-1pt]{$W_{a_2}$};
\draw[Wedge] (v1)--(v2) node[lab,midway,above left=-1pt]{$W_{a_3}$};
\node[ext,label={[lab]above:$x_1$}] at (v1){};
\node[ext,label={[lab]below:$x_2$}] at (v2){};
\node[ext,label={[lab]below:$x_3$}] at (v3){};
\end{scope}
\end{tikzpicture}
}
\caption{The star-triangle relation~\eqref{eq:str}. Open circles denote external spins, the filled circle denotes the internal spin $x_0$, which is integrated over $\Qp$. Dashed and solid lines denote the weights $\overline W$ and $W$, respectively}
\label{fig:str}
\end{figure}

\paragraph{Normalization} The spin-independent factor $R$ depends on the normalization of the weights. We use the inversion relations~\cite{Bazhanov:2016ajm} and fix the remaining scalar freedom by prescribing the Fourier multiplier of the convolution weight. The weight~\eqref{eq:W} obviously satisfies the first inversion relation,
\begin{equation}\label{eq:inv1}
 W_a(x,y)\,W_{-a}(x,y)=1.
\end{equation}
In the second inversion relation the product is replaced by the convolution over the common spin. Using the Fourier transform~\eqref{eq:fourier}, one finds
\begin{equation}\label{eq:inv-raw}
 \int_{\Qp}W_{1-a}(x,z)\,W_{1+a}(z,y)\,dz=\Gamma_p(a)\,\Gamma_p(-a)\,\delta(x-y).
\end{equation}
Therefore it is natural to normalize the star weights by the local gamma function and to introduce
\begin{equation}\label{eq:weights}
 \overline W_a(x,y)=\frac{W_{1-a}(x,y)}{\Gamma_p(a)}=\frac{|x-y|_p^{a-1}}{\Gamma_p(a)}.
\end{equation}
With this normalization both inversion relations hold,
\begin{equation}\label{eq:inv2}
 W_a(x,y)\,W_{-a}(x,y)=1,\qquad
 \int_{\Qp}\overline W_a(x,z)\,\overline W_{-a}(z,y)\,dz=\delta(x-y).
\end{equation}
Since the factor~\eqref{eq:R-factor} is a product of functions of a single parameter, the same normalization removes it from the star-triangle relation. Dividing~\eqref{eq:str-R} by $R(a_1,a_2,a_3)$, one ends up with
\begin{equation}\label{eq:str}
 \int_{\Qp}\prod_{j=1}^3\overline W_{a_j}(x_j,x_0)\,dx_0
 =W_{a_1}(x_2,x_3)\,W_{a_2}(x_1,x_3)\,W_{a_3}(x_1,x_2),
 \qquad \sum_j a_j=1.
\end{equation}
The choice~\eqref{eq:weights} gives $R=1$. Note that the inversion relations alone do not fix this normalization. For example, replacing $\overline W_a$ by $p^a\overline W_a$ preserves the convolution inversion but multiplies the star side of~\eqref{eq:str} by $p$, since $\sum_j a_j=1$. The star-triangle relation~\eqref{eq:str} is shown pictorially in Fig.~\ref{fig:str}. Note also that, by~\eqref{eq:fourier}, the Fourier transform of $\overline W_a(0,x)$ is $|k|_p^{-a}$. Hence the second relation in~\eqref{eq:inv2} is simply $|k|_p^{-a}|k|_p^{a}=1$ in momentum space. We will meet this relation again in Section~\ref{sec:yb} in the form $M(a)M(-a)=1$.

\paragraph{Periodicity} Let $\tau=2\pi i/\ln p$. Since $|x|_p^{\tau}=1$ for $x\ne0$, away from coincident points all exponents are defined modulo $\tau$. In particular,
\begin{equation}\label{eq:eta}
 \eta(x)=(-1)^{v_p(x)}=|x|_p^{\tau/2},\qquad x\ne0,
\end{equation}
is the unramified quadratic character of $\Qp^\times$. This character appears in the exponents of the finite model of Section~\ref{sec:finite}.

\section{Factorized Yang-Baxter operator}\label{sec:yb}

In this section, we construct the Yang-Baxter operator from the star-triangle relation~\eqref{eq:str}. We follow the operator approach of~\cite{Gahramanov:2015cva}, see also~\cite{Gahramanov:2022jxz,Derkachov:2006fw,Chicherin:2012yn}.

\subsection{Star-triangle relation in operator form}\label{sec:opform}

Let us introduce the integral operator $M(a)$ and the multiplication operator $D_{ij}(a)$,
\begin{equation}\label{eq:MD}
 (M(a)f)(x)=\int_{\Qp}\frac{|x-z|_p^{a-1}}{\Gamma_p(a)}\,f(z)\,dz,
 \qquad D_{ij}(a)f=|x_i-x_j|_p^{-a}\,f.
\end{equation}
By the Fourier transform~\eqref{eq:fourier}, the operator $M(a)$ acts as multiplication by $|k|_p^{-a}$ in momentum space. Therefore one has
\begin{equation}\label{eq:Mgroup}
 M(a)M(b)=M(a+b),\qquad M(0)=1,
\end{equation}
and in particular $M(a)M(-a)=1$. Obviously, $D(a)D(-a)=1$ as well.

Let us specify in what sense the operators $M(a)$ should be understood. Let $f$ be a locally constant function with compact support. For $\operatorname{Re}a>0$ the power kernel in~\eqref{eq:MD} is locally integrable. The normalized expression also requires $\Gamma_p(a)\ne0$. On the imaginary axis we use the Fourier multiplier instead. By the Fourier transform~\eqref{eq:fourier}, for $0<\operatorname{Re}a<1$ the integral~\eqref{eq:MD} can be written as
\begin{equation}\label{eq:Mfourier}
 (M(a)f)(x)=\int_{\Qp}|k|_p^{-a}\,\hat f(k)\,\psi(-kx)\,dk,\qquad \hat f(k)=\int_{\Qp}f(z)\,\psi(kz)\,dz .
\end{equation}
The integral~\eqref{eq:Mfourier} converges for all $\operatorname{Re}a<1$ and gives the meromorphic continuation of~\eqref{eq:MD} to this region (see Appendix~\ref{app:unitary}). Therefore we take~\eqref{eq:Mfourier} as the definition of $M(a)$. In particular, one has $M(0)=1$. If moreover $\int_{\Qp}f\,dx=0$, then $\hat f$ vanishes near $k=0$, and~\eqref{eq:Mfourier} is defined for all $a\in\mathbb C$. On such functions the relations~\eqref{eq:Mgroup} hold for all $a$ and $b$. Finally, for $a\in i\mathbb R$ the factors $|k|_p^{-a}$ and $|x_i-x_j|_p^{-a}$ have absolute value $1$, so that $M(a)$ and $D_{ij}(a)$ are unitary operators on $L^2$. This is the case which we need below.

Substituting $(a_1,a_2,a_3)=(s,1-s-t,t)$ in~\eqref{eq:str} and using the reflection formula~\eqref{eq:reflection} for $\Gamma_p(1-s-t)$, one finds
\begin{equation}\label{eq:opstr}
 M(s)\,D(s+t)\,M(t)=D(t)\,M(s+t)\,D(s),
\end{equation}
where $D(a)$ denotes multiplication by $|x-y|_p^{-a}$ with $y$ fixed. In order to derive~\eqref{eq:opstr}, note that the integral kernel of the left-hand side is the integral in~\eqref{eq:rawstr} with $x_0=z$, $x_1=x$, $x_2=y$ and $x_3=w$, divided by $\Gamma_p(s)\Gamma_p(t)$. The right-hand side of~\eqref{eq:rawstr} then gives $|x-y|_p^{-t}\,|x-w|_p^{s+t-1}\Gamma_p(s+t)^{-1}\,|w-y|_p^{-s}$, which is the kernel of the right-hand side of~\eqref{eq:opstr}. The relation~\eqref{eq:opstr} has the same operator structure as the star-triangle relations used in the Bailey-pair construction~\cite{Gahramanov:2015cva,Gahramanov:2022jxz} and in the construction of $SL(2,\mathbb C)$ and conformal R-operators~\cite{Derkachov:2006fw,Chicherin:2012yn}.

Note that the calculation above uses the integral~\eqref{eq:str}, which converges for $\operatorname{Re}s>0$, $\operatorname{Re}t>0$ and $\operatorname{Re}(s+t)<1$. However, for imaginary $s$ and $t$, which we need below, this integral diverges, and the relation~\eqref{eq:opstr} requires a separate proof. We first prove it for matrix elements between suitable test functions, show that these matrix elements are analytic in $s$ and $t$, and then continue them analytically to imaginary values. The precise statement is the following.

\begin{lemma}\label{lem:opstr}
Let $y\in\Qp$, and let $\Phi_y$ be the space of locally constant functions on $\Qp$ with compact support which vanish in a neighborhood of $y$ and have vanishing integral. Then the following statements hold.
\begin{itemize}
 \item[(i)] For all $f,g\in\Phi_y$ and $\operatorname{Re}(s+t)<1$,
 \begin{equation}\label{eq:opstr-matrix}
 \int_{\Qp}\bigl(M(s)f\bigr)(x)\,|x-y|_p^{-s-t}\,\bigl(M(t)g\bigr)(x)\,dx
 =\int_{\Qp}|x-y|_p^{-t}f(x)\,\Bigl(M(s+t)\bigl[\,|\cdot-y|_p^{-s}g\,\bigr]\Bigr)(x)\,dx,
 \end{equation}
 and both sides are analytic functions of $(s,t)$ in this domain.
 \item[(ii)] For $s,t\in i\mathbb R$ the relation~\eqref{eq:opstr} holds as an identity between unitary operators on $L^2(\Qp)$.
\end{itemize}
\end{lemma}

The proof of Lemma~\ref{lem:opstr} is technical, and we present it in Appendix~\ref{app:unitary}.

\subsection{Coxeter relations and factorization}\label{sec:coxeter}

Let us consider functions $f(x_1,x_2)$ of two spins, each carrying a pair of parameters. For our later purposes it is convenient to use the following parametrization,
\begin{equation}\label{eq:lambdas}
 (\lambda_1,\lambda_2,\lambda_3,\lambda_4)
 =(u+g_1,\,u-g_1,\,v+g_2,\,v-g_2),\qquad \delta=u-v.
\end{equation}
Let $s_1,s_2,s_3$ be the elementary transpositions of the permutation group $S_4$ acting on~\eqref{eq:lambdas}. We associate with them the operators
\begin{equation}\label{eq:Si}
 S_1=M_1(\lambda_1-\lambda_2),\qquad
 S_2=D_{12}(\lambda_2-\lambda_3),\qquad
 S_3=M_2(\lambda_3-\lambda_4),
\end{equation}
where $M_1$ and $M_2$ act on $x_1$ and $x_2$, respectively. In other words, each operator depends on the difference of the two parameters which are exchanged. In order to check the Coxeter relations, we take $\delta,g_1,g_2\in i\mathbb R$. In this case all differences $\lambda_i-\lambda_j$ are imaginary, and all operators~\eqref{eq:Si} are unitary on $L^2(\Qp^2)$.
\begin{itemize}
 \item \emph{Inversion.} Since $s_i^2=1$, one needs $S_i(\lambda)\,S_i(s_i\lambda)=1$. This follows from $M(a)M(-a)=1$ and $D(a)D(-a)=1$.
 \item \emph{Commutativity.} The operators $S_1$ and $S_3$ act on different variables and commute.
 \item \emph{Braid relations.} Acting from right to left, the words $s_1s_2s_1$ and $s_2s_1s_2$ give the operators
 \[
 M_1(\lambda_2-\lambda_3)\,D_{12}(\lambda_1-\lambda_3)\,M_1(\lambda_1-\lambda_2)\quad\text{and}\quad D_{12}(\lambda_1-\lambda_2)\,M_1(\lambda_1-\lambda_3)\,D_{12}(\lambda_2-\lambda_3),
 \]
 respectively. With $s=\lambda_2-\lambda_3$ and $t=\lambda_1-\lambda_2$ these are the two sides of the star-triangle relation~\eqref{eq:opstr} in the variable $x_1$, with $y=x_2$ fixed. For almost every $x_2$ they act on $f(\cdot,x_2)\in L^2(\Qp)$ as the two sides of~\eqref{eq:opstr}, and their equality follows from Lemma~\ref{lem:opstr}(ii). The relation $s_2s_3s_2=s_3s_2s_3$ follows in the same way, with $M_2$ instead of $M_1$ and $y=x_1$.
\end{itemize}
Therefore~\eqref{eq:Si} gives a representation of the Coxeter relations of the permutation group of the parameters.

\begin{remark}\label{rem:continuation}
For a product of operators, continuation refers to its matrix elements on fixed test functions. Let $z\in\mathbb C^m$ denote the independent parameters. Suppose the matrix elements of both full products admit meromorphic continuations to a common connected domain $\Omega$, and are holomorphic near a nonempty relatively open subset of $i\mathbb R^m$ contained in $\Omega$. If they agree there with the unitary matrix elements, uniqueness of analytic continuation gives equality throughout the domain as meromorphic functions. Appendix~\ref{app:unitary} constructs such continuations for~\eqref{eq:opstr-matrix} on $\Phi_y$. Existence for longer products requires a separate argument.
\end{remark}

\begin{figure}[htbp]
\centering
{\sffamily
\begin{tikzpicture}
\begin{scope}
\coordinate (x1) at (0,2.2); \coordinate (x2) at (2.6,2.2);
\coordinate (z1) at (0,0);   \coordinate (z2) at (2.6,0);
\draw[Wedge] (x1)--(x2) node[lab,midway,above]{$W_A$};
\draw[Wedge] (z1)--(z2) node[lab,midway,below]{$W_E$};
\draw[Wbar] (x1)--(z1) node[lab,midway,left]{$\overline W_B$};
\draw[Wbar] (x2)--(z2) node[lab,midway,right]{$\overline W_C$};
\node[ext,label={[lab]above:$x_1$}] at (x1){};
\node[ext,label={[lab]above:$x_2$}] at (x2){};
\node[int,label={[lab]below:$z_1$}] at (z1){};
\node[int,label={[lab]below:$z_2$}] at (z2){};
\node[lab] at (1.3,-1.6){(a)};
\end{scope}
\begin{scope}[xshift=5.2cm,yshift=-0.35cm,x=0.75cm,y=0.72cm]
\def\pos{{0,1,2,3}}
\draw[line width=.8pt] (0,0)--(0,1)--(0,2)--(1,3)--(2,4);          
\draw[line width=.8pt] (1,0)--(2,1)--(3,2)--(3,3)--(3,4);          
\draw[line width=.8pt] (2,0)--(1,1)--(1,2)--(0,3)--(0,4);          
\draw[line width=.8pt] (3,0)--(3,1)--(2,2)--(2,3)--(1,4);          
\foreach \j in {1,2,3,4}{\node[lab,below] at (\j-1,0){$\lambda_{\j}$};}
\node[lab,above] at (0,4){$\lambda_3$}; \node[lab,above] at (1,4){$\lambda_4$};
\node[lab,above] at (2,4){$\lambda_1$}; \node[lab,above] at (3,4){$\lambda_2$};
\node[lab,anchor=west] at (3.45,0.5){$s_2:\ D_{12}(E)$};
\node[lab,anchor=west] at (3.45,1.5){$s_3:\ M_2(C)$};
\node[lab,anchor=west] at (3.45,2.5){$s_1:\ M_1(B)$};
\node[lab,anchor=west] at (3.45,3.5){$s_2:\ D_{12}(A)$};
\draw[decorate,decoration={brace,mirror,amplitude=3pt}] (-0.15,-0.62)--(1.15,-0.62) node[lab,midway,below=3pt]{site 1};
\draw[decorate,decoration={brace,mirror,amplitude=3pt}] (1.85,-0.62)--(3.15,-0.62) node[lab,midway,below=3pt]{site 2};
\node[lab] at (2.2,-1.74){(b)};
\end{scope}
\end{tikzpicture}
}
\caption{(a) The integral kernel~\eqref{eq:Rkernel} of the Yang-Baxter operator as a product of four Boltzmann weights. The variables $z_1,z_2$ (filled circles) are integrated over. (b) The word $s_2s_1s_3s_2$ which exchanges the pairs $(\lambda_1,\lambda_2)$ and $(\lambda_3,\lambda_4)$, read from bottom to top, together with the corresponding factors of~\eqref{eq:R}}
\label{fig:R}
\end{figure}
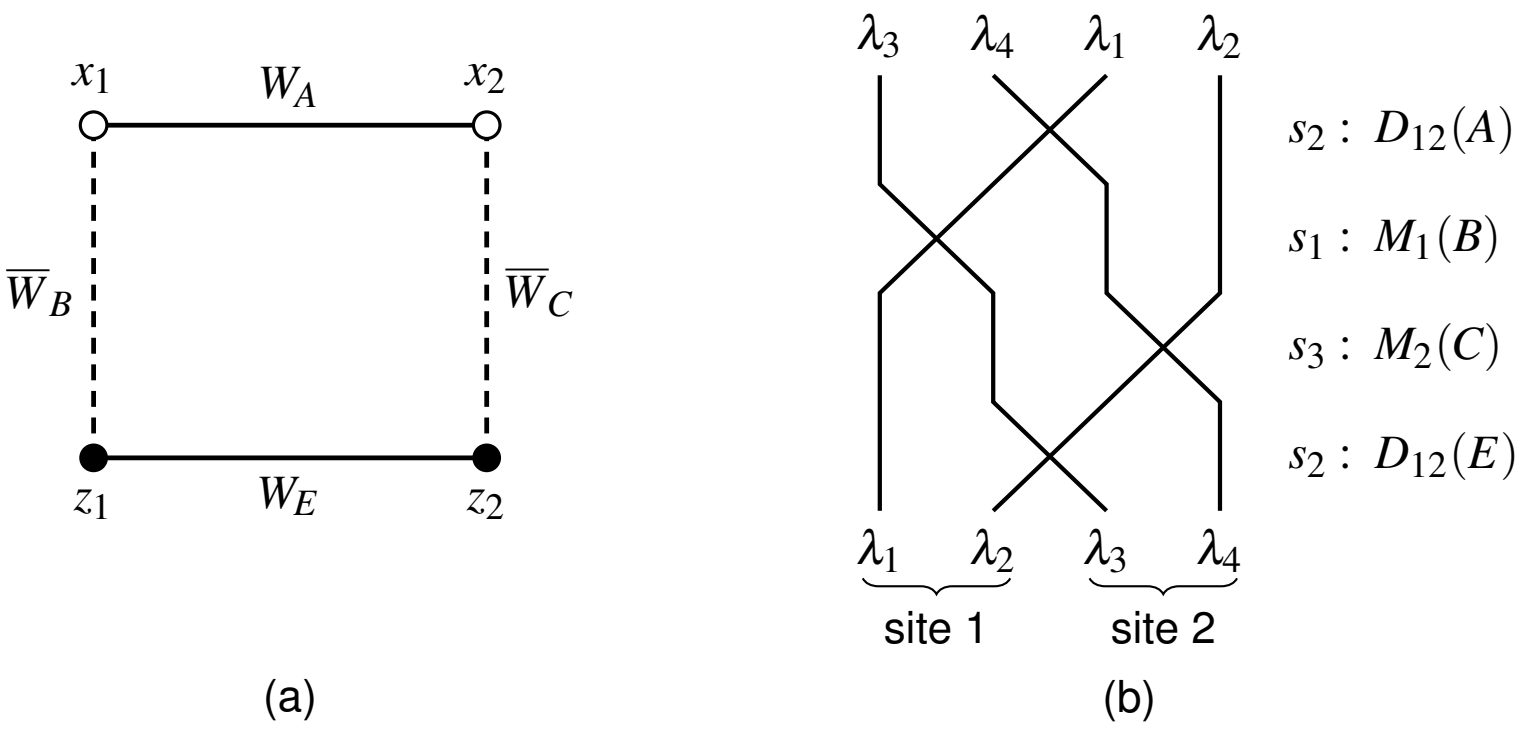

We can now construct the Yang-Baxter operator. It corresponds to the permutation which exchanges the two pairs, $(\lambda_1,\lambda_2)\leftrightarrow(\lambda_3,\lambda_4)$. This permutation is given by the word $s_2s_1s_3s_2$, see Fig.~\ref{fig:R}b. Acting from right to left, the parameters change as
\begin{equation}\label{eq:word}
 (\lambda_1,\lambda_2,\lambda_3,\lambda_4)\xrightarrow{\ s_2\ }
 (\lambda_1,\lambda_3,\lambda_2,\lambda_4)\xrightarrow{\ s_3\ }
 (\lambda_1,\lambda_3,\lambda_4,\lambda_2)\xrightarrow{\ s_1\ }
 (\lambda_3,\lambda_1,\lambda_4,\lambda_2)\xrightarrow{\ s_2\ }
 (\lambda_3,\lambda_4,\lambda_1,\lambda_2).
\end{equation}
Collecting the corresponding operators, we obtain
\begin{equation}\label{eq:R}
 \Rc_{12}(\delta;g_1,g_2)=D_{12}(A)\,M_1(B)\,M_2(C)\,D_{12}(E),
\end{equation}
where
\begin{equation}\label{eq:ABCE}
 A=\delta+g_1+g_2,\qquad B=\delta+g_1-g_2,\qquad
 C=\delta-g_1+g_2,\qquad E=\delta-g_1-g_2.
\end{equation}
The integral kernel of~\eqref{eq:R} has the following form
\begin{equation}\label{eq:Rkernel}
 \Rc(x_1,x_2;z_1,z_2)=
 \frac{|x_{12}|_p^{-A}\,|x_1-z_1|_p^{B-1}\,|x_2-z_2|_p^{C-1}\,|z_{12}|_p^{-E}}
 {\Gamma_p(B)\,\Gamma_p(C)},
\end{equation}
which is~\eqref{eq:intro-R}. The kernel is a product of four Boltzmann weights~\eqref{eq:W} and~\eqref{eq:weights}, namely two weights $W$ along the edges $x_1x_2$ and $z_1z_2$, and two weights $\overline W$ along the edges $x_1z_1$ and $x_2z_2$, see Fig.~\ref{fig:R}a. Note that for $\operatorname{Re}B\le0$ or $\operatorname{Re}C\le0$ the factors $|x_1-z_1|_p^{B-1}/\Gamma_p(B)$ and $|x_2-z_2|_p^{C-1}/\Gamma_p(C)$ should be understood as the distribution kernels of the operators $M_1(B)$ and $M_2(C)$. On the unitary locus, the ordered product~\eqref{eq:R} specifies the full distribution kernel across the diagonals.

\subsection{Intertwiner of principal series representations}\label{sec:intertwining}

In this subsection, we show that the operator~\eqref{eq:R} intertwines the Hilbert tensor product of two unitary principal series representations of $\GL$ with the same product in the opposite order. Here we briefly outline some basic facts and refer the reader to~\cite{Gelfand:1969} for more details. For $h=\left(\begin{smallmatrix}\mathsf a&\mathsf b\\ \mathsf c&\mathsf d\end{smallmatrix}\right)\in\GL$ and $\rho\in\mathbb C$ we define
\begin{equation}\label{eq:Trho}
 \bigl(T_\rho(h)f\bigr)(x)=|\det h|_p^{\rho}\,|\mathsf cx+\mathsf d|_p^{-2\rho}\,f(h\cdot x),
 \qquad h\cdot x=\frac{\mathsf ax+\mathsf b}{\mathsf cx+\mathsf d}.
\end{equation}
One has $T_\rho(h_1)T_\rho(h_2)=T_\rho(h_2h_1)$, so $h\mapsto T_\rho(h^{-1})$ is a representation.\footnote{The center of $\GL$ acts trivially, so~\eqref{eq:Trho} is in fact a representation of $PGL_2(\Qp)$.} The operators $T_\rho(h)$ act on the space $V_\rho$ of locally constant functions $f$ on $\Qp$ such that $f(x)=C\,|x|_p^{-2\rho}$ for large $|x|_p$. This behavior at infinity is created by the inversion $h\cdot x=1/x$, for which $(T_\rho(h)f)(x)=|x|_p^{-2\rho}f(1/x)$. The representation~\eqref{eq:Trho} is the principal series representation of $\GL$ induced from the character $|\cdot|_p^{\rho}$. For $\operatorname{Re}\rho=\tfrac12$, i.e.\ for $\rho=\tfrac12+g$ with $g\in i\mathbb R$, it is unitary with respect to the $L^2$ norm and extends to a unitary representation on $L^2(\Qp)$ (the unitary principal series). We write $\mathcal H_\rho$ for this completion of $V_\rho$ and $\widehat\otimes$ for the Hilbert tensor product. On functions of two variables we use
\begin{equation}
 \bigl((T_{\rho_1}\otimes T_{\rho_2})(h)f\bigr)(x_1,x_2)=|\det h|_p^{\rho_1+\rho_2}\,|\mathsf cx_1+\mathsf d|_p^{-2\rho_1}\,|\mathsf cx_2+\mathsf d|_p^{-2\rho_2}\,f(h\cdot x_1,h\cdot x_2).
\end{equation}
We say that an operator $A$ intertwines $T$ with $T'$ if $A\,T(h)=T'(h)\,A$ for all $h\in\GL$.

\begin{lemma}\label{lem:intertwiners}
For all $h\in\GL$,
\begin{align}
 M(a)\,T_{\frac{1+a}{2}}(h)&=T_{\frac{1-a}{2}}(h)\,M(a),\label{eq:Mint}\\
 D_{12}(a)\,\bigl(T_{\rho_1}\otimes T_{\rho_2}\bigr)(h)&=\bigl(T_{\rho_1+\frac a2}\otimes T_{\rho_2+\frac a2}\bigr)(h)\,D_{12}(a).\label{eq:Dint}
\end{align}
The relation~\eqref{eq:Dint} is a pointwise identity away from the diagonals and poles of the coordinate maps, for all $a\in\mathbb C$. The relation~\eqref{eq:Mint} holds for $0<\operatorname{Re}a<1$ on $V_{(1+a)/2}$, with $M(a)$ given by the convergent integral~\eqref{eq:MD}, and for $a\in i\mathbb R$ as an identity between unitary operators on $L^2(\Qp)$.
\end{lemma}

For $a=2g$ on the unitary axis it states that $M(2g)$ intertwines $T_{\frac12+g}$ with $T_{\frac12-g}$, i.e.\ $M(2g)$ is the standard intertwining operator of the principal series~\cite{Gelfand:1969}. The proof of Lemma~\ref{lem:intertwiners} is based on a simple change of variables, and we present it in Appendix~\ref{app:proof}.

Applying Lemma~\ref{lem:intertwiners} to each factor of the factorization~\eqref{eq:R}, we obtain the following statement, which is part (ii) of Theorem~\ref{thm:main}.

\begin{proposition}\label{prop:intertwiner}
Let $\rho_j=\tfrac12+g_j$ with $\delta,g_1,g_2\in i\mathbb R$. Then for all $h\in\GL$,
\begin{equation}\label{eq:Rint}
 \Rc_{12}(\delta;g_1,g_2)\,\bigl(T_{\rho_1}\otimes T_{\rho_2}\bigr)(h)
 =\bigl(T_{\rho_2}\otimes T_{\rho_1}\bigr)(h)\,\Rc_{12}(\delta;g_1,g_2)
\end{equation}
as an identity between unitary operators on $L^2(\Qp^2)$. In particular, for $g_1=g_2=g$ the operator $\Rc(\delta;g,g)$ commutes with the diagonal action of $\GL$ on $\mathcal H_{\frac12+g}\widehat\otimes\mathcal H_{\frac12+g}$.
\end{proposition}

\begin{proof}
We follow the factors of~\eqref{eq:R} from right to left. By~\eqref{eq:Dint} with $a=E$, the operator $D_{12}(E)$ intertwines $T_{\frac12+g_1}\otimes T_{\frac12+g_2}$ with $T_{\frac12(1+B)}\otimes T_{\frac12(1+C)}$, since $\tfrac12+g_1+\tfrac E2=\tfrac12(1+B)$ and $\tfrac12+g_2+\tfrac E2=\tfrac12(1+C)$. Since $M_1$ and $M_2$ act on one variable each,~\eqref{eq:Mint} shows that $M_1(B)M_2(C)$ intertwines the latter with $T_{\frac12(1-B)}\otimes T_{\frac12(1-C)}$. Finally, by~\eqref{eq:Dint} with $a=A$, the operator $D_{12}(A)$ intertwines this with $T_{\frac12+g_2}\otimes T_{\frac12+g_1}$, since $\tfrac12(1-B)+\tfrac A2=\tfrac12+g_2$ and $\tfrac12(1-C)+\tfrac A2=\tfrac12+g_1$. The exponents $A,B,C,E$ are imaginary and all representation labels have real part $1/2$, so each factor is unitary and the composition of the three relations gives~\eqref{eq:Rint}.
\end{proof}
The following example with characteristic functions of balls shows that the Hilbert completion is needed. Take $g_1=g_2=0$, $\delta\in i\mathbb R$ with $p^\delta\ne1$, and disjoint balls $U,V,B_0$ of equal radius. Apply~\eqref{eq:R} to $\mathbf 1_U\otimes\mathbf 1_V$. On $U\times V$ the rightmost multiplication factor is constant, and on $B_0\times U$ and $B_0\times V$ the respective convolution kernels are nonzero constants. Thus on $B_0^2$ off the diagonal the result is $c\,|x_1-x_2|_p^{-\delta}$ with a constant $c\ne0$. Every neighborhood of a diagonal point contains shells of positive measure on which this function takes different values. Hence the result has no locally constant representative there, whereas every element of $V_{1/2}\otimes V_{1/2}$ has one.

The representations depend only on the differences inside each pair of parameters, i.e.\ on the spin parameters $g_j$, while the spectral parameter $\delta$ is not a representation label. This is different from the $SL(2,\mathbb C)$ case considered by Derkachov and Manashov~\cite{Derkachov:2006fw}, where each site carries two parameters, one for the holomorphic and one for the antiholomorphic part. The spin parameter $g_j\in i\mathbb R$ corresponds to the unitary principal series.

\subsection{Yang-Baxter equation, initial condition and unitarity}\label{sec:properties}

In this subsection, we discuss the properties of the Yang-Baxter operator~\eqref{eq:R}. We start with the Yang-Baxter equation. As in~\cite{Gahramanov:2015cva,Derkachov:2006fw}, it follows from the Coxeter relations of the permutation group. For three sites we use the following parameters,
\begin{equation}\label{eq:lambdas6}
 (\lambda_1,\ldots,\lambda_6)=(u_1+g_1,\,u_1-g_1,\,u_2+g_2,\,u_2-g_2,\,u_3+g_3,\,u_3-g_3),
\end{equation}
and represent the elementary transpositions $s_1,\ldots,s_5$ of the permutation group $S_6$ by the operators
\begin{equation}\label{eq:Si6}
 S_1=M_1(\lambda_1-\lambda_2),\quad S_2=D_{12}(\lambda_2-\lambda_3),\quad S_3=M_2(\lambda_3-\lambda_4),\quad
 S_4=D_{23}(\lambda_4-\lambda_5),\quad S_5=M_3(\lambda_5-\lambda_6).
\end{equation}
In addition to the relations checked in Section~\ref{sec:coxeter}, one needs the commutativity of $S_i$ and $S_j$ for $|i-j|>1$ and the braid relations for the pairs $(S_3,S_4)$ and $(S_4,S_5)$. The commutativity is obvious, since in each case either both operators are multiplication operators or they act on different variables. The two braid relations are again the star-triangle relation~\eqref{eq:opstr}, written for the variable $x_2$ with $y=x_3$ and for the variable $x_3$ with $y=x_2$, respectively. Therefore~\eqref{eq:Si6} gives a representation of the Coxeter relations of $S_6$, and the product of the operators along a reduced word depends only on the permutation.\footnote{This follows from the theorem of Matsumoto, which states that any two reduced words of a permutation are related by braid and commutation moves.} As before, for $g_j\in i\mathbb R$ and $u_j-u_k\in i\mathbb R$ all differences $\lambda_i-\lambda_j$ are imaginary, all these operators are unitary on $L^2(\Qp^3)$, and the relations hold as identities between unitary operators.

The operators $\Rc_{12}$ and $\Rc_{23}$ correspond to the words $s_2s_1s_3s_2$ and $s_4s_3s_5s_4$, i.e.\ they exchange two neighboring pairs of parameters. Both sides of the equation below are reduced words of length $12$ for the permutation which reverses the order of the three pairs. As a result, one finds
\begin{equation}\label{eq:ybe-general}
\begin{split}
 \Rc_{23}(u_1-u_2;g_1,g_2)\,&\Rc_{12}(u_1-u_3;g_1,g_3)\,\Rc_{23}(u_2-u_3;g_2,g_3)\\
 &=\Rc_{12}(u_2-u_3;g_2,g_3)\,\Rc_{23}(u_1-u_3;g_1,g_3)\,\Rc_{12}(u_1-u_2;g_1,g_2).
\end{split}
\end{equation}
For $g_j\in i\mathbb R$ and $u_j-u_k\in i\mathbb R$ both sides are products of twelve unitary operators, and~\eqref{eq:ybe-general} is an identity between unitary operators on $L^2(\Qp^3)$. This proves part (i) of Theorem~\ref{thm:main}. By Proposition~\ref{prop:intertwiner}, both sides intertwine the diagonal actions on $\mathcal H_{\rho_1}\widehat\otimes\mathcal H_{\rho_2}\widehat\otimes\mathcal H_{\rho_3}$ and $\mathcal H_{\rho_3}\widehat\otimes\mathcal H_{\rho_2}\widehat\otimes\mathcal H_{\rho_1}$, where $\rho_j=\tfrac12+g_j$. Outside the unitary locus, equations~\eqref{eq:Rint} and~\eqref{eq:ybe-general} extend wherever the matrix elements of the full products satisfy the hypotheses of Remark~\ref{rem:continuation}. For identical spin parameters,~\eqref{eq:ybe-general} reduces to the braid form
\begin{equation}\label{eq:ybe}
 \Rc_{23}(u_1-u_2)\,\Rc_{12}(u_1-u_3)\,\Rc_{23}(u_2-u_3)
 =\Rc_{12}(u_2-u_3)\,\Rc_{23}(u_1-u_3)\,\Rc_{12}(u_1-u_2).
\end{equation}
On the same unitary locus the factorization gives
\begin{equation}\label{eq:unitarity-general}
 \Rc(\delta;g_1,g_2)\,\Rc(-\delta;g_2,g_1)=1,
\end{equation}
since the four factors of $\Rc(-\delta;g_2,g_1)$ have the exponents $-E,-B,-C,-A$ and cancel those of $\Rc(\delta;g_1,g_2)$ pairwise. Moreover, for $\delta,g_1,g_2\in i\mathbb R$ the adjoint operators are $M(a)^\dagger=M(-a)$ and $D(a)^\dagger=D(-a)$, so that $\Rc(\delta;g_1,g_2)^\dagger=\Rc(-\delta;g_2,g_1)$. Hence in this case $\Rc(\delta;g_1,g_2)$ is a unitary operator on $L^2(\Qp^2)$ in the usual sense, which is part (iii) of Theorem~\ref{thm:main}. For identical spin parameters one has the initial condition and unitarity in the usual form,
\begin{equation}\label{eq:unitarity}
 \Rc(0)=1,\qquad \Rc(\delta)\,\Rc(-\delta)=1.
\end{equation}
Indeed, at $\delta=0$ one has $B=C=0$, and $\Rc(0)=D_{12}(2g)D_{12}(-2g)=1$, which is part (iv) of Theorem~\ref{thm:main}. For $g_1\ne g_2$ the operator
\[
 \Rc(0;g_1,g_2)=D_{12}(g_1+g_2)\,M_1(g_1-g_2)\,M_2(g_2-g_1)\,D_{12}(-g_1-g_2)
\]
is generically nontrivial and intertwines the diagonal actions on $\mathcal H_{\frac12+g_1}\widehat\otimes\mathcal H_{\frac12+g_2}$ and $\mathcal H_{\frac12+g_2}\widehat\otimes\mathcal H_{\frac12+g_1}$.

Finally, for identical spins one has $B=C=\delta$. In this case, for generic $g$, the scalar kernel away from its diagonals has a double pole at $\delta=1$, and its leading coefficient factorizes. We discuss this coefficient and the composition of the resulting singular kernels in Appendix~\ref{app:crossing}.

\section{Bailey lemma and four-point integrals}\label{sec:bailey}

In this section, we show that the operator form~\eqref{eq:opstr} of the star-triangle relation leads to a Bailey lemma. We follow the integral Bailey lemma introduced in~\cite{Gahramanov:2015cva}, see also its operator form in~\cite{Gahramanov:2022jxz}. As in the construction of the Yang-Baxter operator, the only input is the relation~\eqref{eq:opstr}.\footnote{The functions $\alpha$ and $\beta$ of this section should not be confused with the constant $\alpha$ of Appendix~\ref{app:wu}.}

\subsection{The Bailey lemma}\label{sec:baileylemma}
We fix a point $y\in\Qp$. As in Section~\ref{sec:opform}, $D(a)$ denotes the multiplication by $|x-y|_p^{-a}$. We say that two functions $\alpha(x)$ and $\beta(x)$ form a Bailey pair with respect to $t$ if $\beta=M(t)\alpha$, i.e.
\begin{equation}\label{eq:baileypair}
\beta(x)=\int_{\Qp}\frac{|x-w|_p^{t-1}}{\Gamma_p(t)}\,\alpha(w)\,dw .
\end{equation}

\begin{lemma}[Bailey lemma]\label{lem:bailey}
Let $(\alpha,\beta)$ be a Bailey pair with respect to $t$. Then the functions
\begin{equation}\label{eq:baileylemma}
\alpha'=D(s)\,\alpha,\qquad \beta'=D(-t)\,M(s)\,D(s+t)\,\beta
\end{equation}
form a Bailey pair with respect to $s+t$, i.e.\ $\beta'=M(s+t)\alpha'$. This holds in each of the following two cases.
\begin{itemize}
\item[(i)] $s,t\in i\mathbb{R}$ and $\alpha\in L^2(\Qp)$.
\item[(ii)] $\operatorname{Re}s>0$, $\operatorname{Re}t>0$, $\operatorname{Re}(s+t)<1$, and for a given $x\neq y$ the integral
\begin{equation}\label{eq:baileycond}
\int_{\Qp}|x-w|_p^{\operatorname{Re}(s+t)-1}\,|w-y|_p^{-\operatorname{Re}s}\,|\alpha(w)|\,dw
\end{equation}
converges. In this case $\beta(z)$ is given by the absolutely convergent integral~\eqref{eq:baileypair} for almost all $z$, the integral which defines $\beta'(x)$ converges absolutely, and $\beta'(x)=\big(M(s+t)\alpha'\big)(x)$.
\end{itemize}
\end{lemma}

\begin{proof}
By the star-triangle relation~\eqref{eq:opstr},
\begin{equation}\label{eq:baileyproof}
\beta'=D(-t)\,M(s)\,D(s+t)\,M(t)\,\alpha=D(-t)\,D(t)\,M(s+t)\,D(s)\,\alpha=M(s+t)\,\alpha' .
\end{equation}
In case (i) all operators in~\eqref{eq:baileyproof} are unitary on $L^2(\Qp)$, and~\eqref{eq:baileyproof} is an identity between unitary operators by Lemma~\ref{lem:opstr}(ii).

In case (ii) we write $\beta'(x)$ as the double integral
\begin{equation}\label{eq:baileydouble}
\beta'(x)=\frac{|x-y|_p^{t}}{\Gamma_p(s)\Gamma_p(t)}\int_{\Qp^2}|x-z|_p^{s-1}\,|z-y|_p^{-s-t}\,|z-w|_p^{t-1}\,\alpha(w)\,dz\,dw .
\end{equation}
Let $\sigma_1=\operatorname{Re}s$ and $\sigma_2=\operatorname{Re}t$. Replacing all functions in~\eqref{eq:baileydouble} by their absolute values and integrating over $z$ by means of~\eqref{eq:rawstr} with the real exponents $(\sigma_1,1-\sigma_1-\sigma_2,\sigma_2)$, one obtains the integral~\eqref{eq:baileycond} multiplied by the finite factor $\Gamma_p(\sigma_1)\Gamma_p(1-\sigma_1-\sigma_2)\Gamma_p(\sigma_2)/|\Gamma_p(s)\Gamma_p(t)|$. Therefore the double integral converges absolutely, and one can integrate in either order. Integrating first over $w$, one obtains $\beta(z)$, which is thus finite for almost all $z$, and then the definition of $\beta'(x)$. Integrating first over $z$ by means of~\eqref{eq:rawstr}, as in the derivation of~\eqref{eq:opstr}, one obtains
\begin{equation*}
\beta'(x)=\int_{\Qp}\frac{|x-w|_p^{s+t-1}}{\Gamma_p(s+t)}\,|w-y|_p^{-s}\,\alpha(w)\,dw=\big(M(s+t)\alpha'\big)(x).
\end{equation*}
\end{proof}

For a given seed, any extension beyond these two cases requires continuation of the expressions on both sides of~\eqref{eq:baileyproof}. For the power-law seed below, this follows from the shell decomposition of the resulting scalar integrals.

Thus the Bailey lemma is the star-triangle relation~\eqref{eq:opstr} written as a map on pairs of functions. The function $\alpha$ is multiplied by a power of $|x-y|_p$, while two multiplication operators and one integral operator act on $\beta$. The derivation~\eqref{eq:baileyproof} uses~\eqref{eq:opstr} in the same way as the braid relation $s_1s_2s_1=s_2s_1s_2$ in Section~\ref{sec:coxeter}.

Taken alone, the lemma contains no more information than the relation~\eqref{eq:opstr}. As in~\cite{Gahramanov:2015cva}, new identities appear when one starts from a particular Bailey pair (a seed) for which $\beta$ can be computed explicitly.

\subsection{A seed and a four-point transformation}\label{sec:baileyseed}
Following~\cite{Gahramanov:2015cva}, let us choose a seed for which $\beta=M(t)\alpha$ is again computed by the star-triangle relation. Let $u_1,u_2$ be two distinct points, different from $y$, and
\begin{equation}\label{eq:seed}
\alpha(w)=|w-u_1|_p^{c_1-1}\,|w-u_2|_p^{c_2-1},\qquad t+c_1+c_2=1 .
\end{equation}
For $\operatorname{Re}t$, $\operatorname{Re}c_1$, $\operatorname{Re}c_2>0$ the integral~\eqref{eq:baileypair} is the integral~\eqref{eq:rawstr} with the points $(x,u_1,u_2)$ and the exponents $(t,c_1,c_2)$. Hence
\begin{equation}\label{eq:seedbeta}
\beta(x)=\Gamma_p(c_1)\,\Gamma_p(c_2)\,|u_1-u_2|_p^{-t}\,|x-u_1|_p^{-c_2}\,|x-u_2|_p^{-c_1},
\end{equation}
and~\eqref{eq:seed},~\eqref{eq:seedbeta} is an explicit Bailey pair. Let us apply the lemma to it. Then both sides of $\beta'=M(s+t)\alpha'$ are integrals of products of four powers. In order to write the result in a symmetric form, we denote $x_1=x$, $x_2=y$, $x_3=u_1$, $x_4=u_2$ and introduce the four-point integral
\begin{equation}\label{eq:fourpoint}
I(a_1,a_2,a_3,a_4)=\int_{\Qp}\prod_{j=1}^4|x_j-x_0|_p^{a_j-1}\,dx_0,\qquad a_1+a_2+a_3+a_4=2 .
\end{equation}
Because of the condition on the exponents, the integrand decreases as $|x_0|_p^{-2}$ at infinity.

\begin{proposition}\label{prop:fourpoint}
Let $x_1,\dots,x_4\in\Qp$ be distinct, $a_1+a_2+a_3+a_4=2$, and
\begin{equation}\label{eq:fourpointcond}
\operatorname{Re}a_j>0\ \ (j=1,\dots,4),\qquad \operatorname{Re}a_3<1,\qquad \operatorname{Re}a_4<1,\qquad \operatorname{Re}(a_1+a_2)<1 .
\end{equation}
Then
\begin{equation}\label{eq:fourpointtrans}
\left|\frac{x_{12}}{x_{34}}\right|_p^{1-a_1-a_2} I(a_1,a_2,a_3,a_4)=\Gamma_p(a_1)\,\Gamma_p(a_2)\,\Gamma_p(a_3)\,\Gamma_p(a_4)\; I(1-a_2,1-a_1,1-a_4,1-a_3).
\end{equation}
\end{proposition}

\begin{proof}
We put
\begin{equation*}
s=a_1,\qquad t=1-a_1-a_2=a_3+a_4-1,\qquad c_1=1-a_4,\qquad c_2=1-a_3 ,
\end{equation*}
so that $t+c_1+c_2=1$. By~\eqref{eq:fourpointcond} one has $\operatorname{Re}s>0$, $\operatorname{Re}t>0$, $\operatorname{Re}(s+t)=\operatorname{Re}(1-a_2)<1$ and $\operatorname{Re}c_1,\operatorname{Re}c_2>0$. Hence~\eqref{eq:seedbeta} holds. The condition~\eqref{eq:baileycond} is satisfied as well. Indeed, the integrand of~\eqref{eq:baileycond} behaves near the four points as powers with the exponents $\operatorname{Re}(s+t)-1$, $-\operatorname{Re}s$, $\operatorname{Re}c_1-1$ and $\operatorname{Re}c_2-1$, respectively, which are all larger than $-1$, and at infinity as $|w|_p^{-2}$. Thus by Lemma~\ref{lem:bailey}(ii) one has $\beta'(x)=(M(s+t)\alpha')(x)$. By~\eqref{eq:seedbeta} the left-hand side is
\begin{equation*}
\beta'(x)=\frac{\Gamma_p(c_1)\Gamma_p(c_2)}{\Gamma_p(s)}\left|\frac{x_{12}}{x_{34}}\right|_p^{t} I(s,1-s-t,1-c_2,1-c_1),
\end{equation*}
while the right-hand side is
\begin{equation*}
\big(M(s+t)\alpha'\big)(x)=\frac{1}{\Gamma_p(s+t)}\,I(s+t,1-s,c_1,c_2).
\end{equation*}
In terms of the exponents $a_j$ one has $(s,1-s-t,1-c_2,1-c_1)=(a_1,a_2,a_3,a_4)$ and $(s+t,1-s,c_1,c_2)=(1-a_2,1-a_1,1-a_4,1-a_3)$. Finally, by the reflection formula~\eqref{eq:reflection}, $\Gamma_p(c_1)=1/\Gamma_p(a_4)$, $\Gamma_p(c_2)=1/\Gamma_p(a_3)$ and $1/\Gamma_p(s+t)=\Gamma_p(a_2)$, which gives~\eqref{eq:fourpointtrans}.
\end{proof}

The transformed exponents $b=(1-a_2,1-a_1,1-a_4,1-a_3)$ also sum to $2$. For fixed distinct $x_j$, both sides of~\eqref{eq:fourpointtrans} are meromorphic in the exponents by the shell decomposition below. The transformation is an involution of this continued identity. Indeed, the second application returns the original exponents, the two distance factors cancel and $\prod_j\Gamma_p(a_j)\Gamma_p(1-a_j)=1$. Continuation is needed for this step, since $\operatorname{Re}(b_1+b_2)>1$ in the chamber~\eqref{eq:fourpointcond}.

All identities obtained in this way are elementary in the following sense. Splitting $\Qp$ into discs which contain at most one of the points, as in the proof of Lemma~\ref{lem:unbalanced}, one reduces any integral of a product of powers $|x_0-x_j|_p^{a_j-1}$ to a finite number of geometric series. Hence~\eqref{eq:fourpointtrans} is an identity between rational functions of $p^{a_j}$ for any fixed configuration of the points. In this respect the situation is different from~\cite{Gahramanov:2015cva,Gahramanov:2022jxz}, where the same construction leads to transformation formulas for hypergeometric integrals.

\subsection{The Bailey chain}\label{sec:baileychain}
Let us now apply the lemma repeatedly, each time to the pair obtained at the previous step. Since the definition~\eqref{eq:baileypair} of a Bailey pair does not involve the point $y$, one may take a different point at each step. We denote by $D_y(a)$ the multiplication by $|x-y|_p^{-a}$. Let $(\alpha,\beta)$ be a Bailey pair with respect to $t_0=t$, and choose $s_k$ and $y_k$, $k=1,\dots,n$. Put $t_k=t_{k-1}+s_k$, $\alpha^{(0)}=\alpha$, $\beta^{(0)}=\beta$, and
\begin{equation}\label{eq:chainstep}
\alpha^{(k)}=D_{y_k}(s_k)\,\alpha^{(k-1)},\qquad \beta^{(k)}=D_{y_k}(-t_{k-1})\,M(s_k)\,D_{y_k}(t_k)\,\beta^{(k-1)} .
\end{equation}
By Lemma~\ref{lem:bailey}, applied $n$ times, $(\alpha^{(n)},\beta^{(n)})$ is a Bailey pair with respect to $t_n$, provided the conditions of the lemma hold at each step. In other words,
\begin{equation}\label{eq:chain}
\beta^{(n)}=M(t_n)\,\alpha^{(n)},\qquad \alpha^{(n)}(x)=\prod_{k=1}^n|x-y_k|_p^{-s_k}\,\alpha(x) .
\end{equation}
The left-hand side of~\eqref{eq:chain} is an $n$-fold integral, while the right-hand side is a single integral. For the seed~\eqref{eq:seed} the latter is an integral of a product of $n+3$ powers, with the points $x,y_1,\dots,y_n,u_1,u_2$, and the powers add up to $t_0+c_1+c_2-3=-2$, as in~\eqref{eq:fourpoint}. For $n=1$ the relation~\eqref{eq:chain} reduces to~\eqref{eq:fourpointtrans}.

\section{Knot invariants from residue-disc averaging}\label{sec:jones}

The star-triangle relation~\eqref{eq:str}, which leads to the Yang-Baxter operator~\eqref{eq:R}, describes the local change of a knot diagram under the third Reidemeister move. Therefore it is natural to use the same Boltzmann weights in order to construct knot invariants. In this section, we use the edge-interaction construction of Wu~\cite{Wu:1992zzb}, which we briefly review in Appendix~\ref{app:wu}.

Let us recall the main ingredients of this construction. One colors the faces of a link diagram as a checkerboard, places a spin with $q$ possible values on each shaded face, and assigns the weight $W_+$ or $W_-$ to each crossing. The weights have to satisfy the inversion relations~\eqref{eq:W2} and~\eqref{eq:W3} for the second Reidemeister move, the star-triangle relation~\eqref{eq:W4} for the third move, and the normalization~\eqref{eq:W1}, which controls the first move. For the Potts weights~\eqref{eq:potts} with $q=t+2+t^{-1}$, the normalized partition function~\eqref{eq:jones} gives the Jones polynomial $V_K(t)$. Below we show that the $p$-adic weights, averaged over residue discs, lead to the Potts weights at $t+t^{-1}=p-1$.

\subsection{Averaging over residue discs and inversion relations}\label{sec:finite}

The continuous power-law weights do not satisfy both inversion relations~\eqref{eq:W2} and~\eqref{eq:W3}, and the plus equation in~\eqref{eq:W4} fails throughout the convergence region of its integral (Proposition~\ref{prop:nostr}). Appendix~\ref{app:continuous} also examines a singular specialization. Therefore we pass to the projective line $\mathbb P^1(\Qp)$ and average over its first residue level in order to obtain a finite spin space. For this purpose we use the chordal version of the weight,
\begin{equation}\label{eq:chordal}
 W_a^{\rm c}(x,y)=[x,y]^{-a},\qquad
 [x,y]=\frac{|x-y|_p}{\max(1,|x|_p)\max(1,|y|_p)},
\end{equation}
where $[x,\infty]=1/\max(1,|x|_p)$. On $\Zp$ the chordal weight agrees with $|x-y|_p^{-a}$. We equip the projective line with the probability measure
\begin{equation}\label{eq:mu}
 d\mu(x)=\frac{p}{p+1}\frac{dx}{\max(1,|x|_p)^2},\qquad
 \int_{\mathbb P^1(\Qp)}d\mu(x)=1.
\end{equation}
The point at infinity has zero measure, and both $[x,y]$ and $\mu$ are invariant under $PGL_2(\Zp)$.

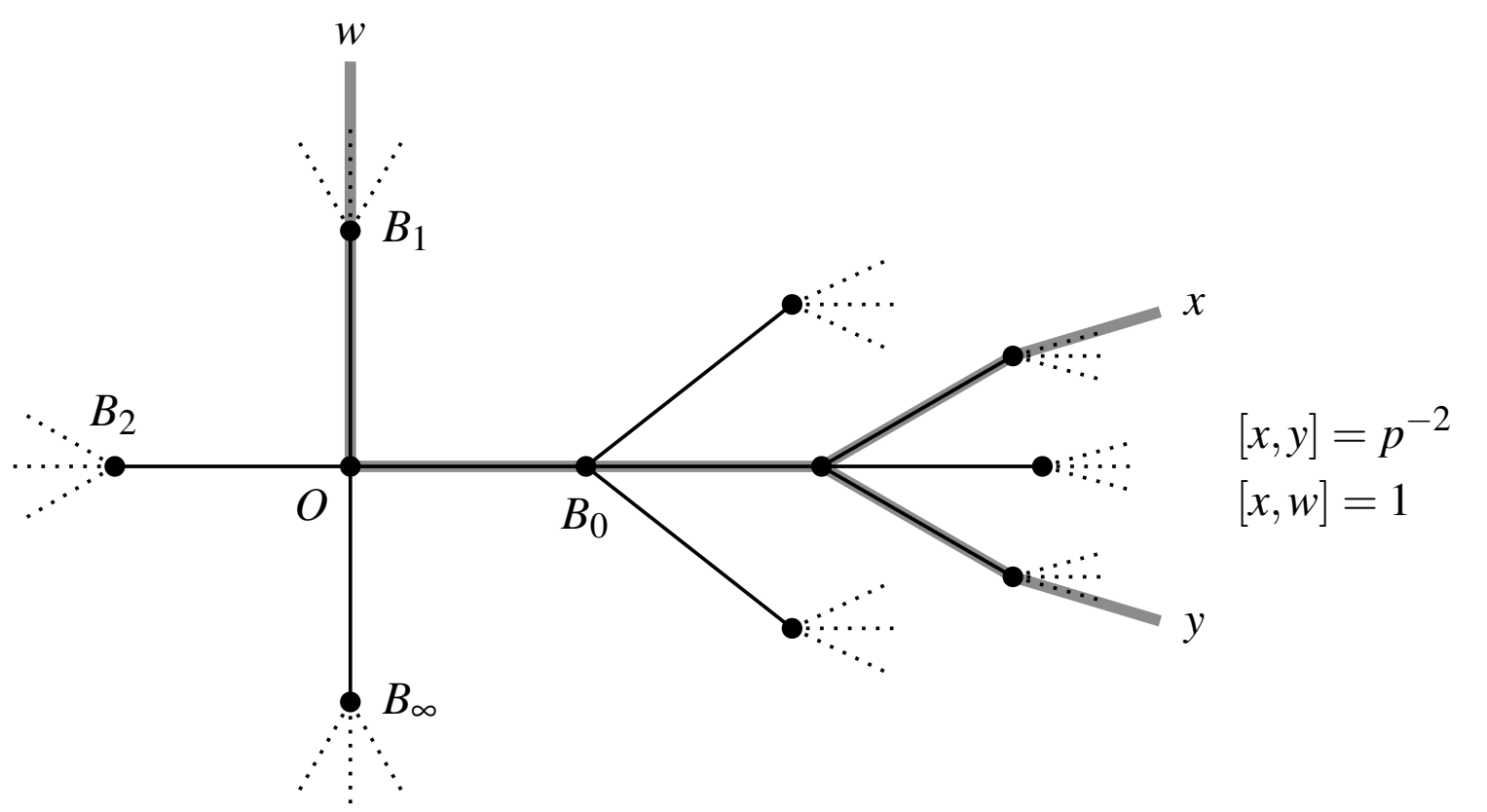
\begin{figure}[htbp]
\centering
{\sffamily
\begin{tikzpicture}[v/.style={circle,fill=black,inner sep=0pt,minimum size=4pt},
  e/.style={line width=.7pt}, path/.style={line width=2.2pt,black!45}]
\coordinate (O) at (0,0);
\coordinate (b0) at (1.6,0);
\coordinate (b1) at (0,1.6);
\coordinate (b2) at (-1.6,0);
\coordinate (bi) at (0,-1.6);
\coordinate (c0) at (3.0,1.1); \coordinate (c1) at (3.2,0); \coordinate (c2) at (3.0,-1.1);
\coordinate (d0) at (4.5,0.75); \coordinate (d1) at (4.7,0); \coordinate (d2) at (4.5,-0.75);
\draw[path] (O)--(b0)--(c1);
\draw[path] (c1)--(d0)--(5.5,1.05);
\draw[path] (c1)--(d2)--(5.5,-1.05);
\draw[path] (O)--(b1)--(0,2.75);
\draw[e] (O)--(b0) (O)--(b1) (O)--(b2) (O)--(bi);
\draw[e] (b0)--(c0) (b0)--(c1) (b0)--(c2);
\draw[e] (c1)--(d0) (c1)--(d1) (c1)--(d2);
\foreach \a/\b in {b1/90,b2/180,bi/270}{
  \draw[e,dotted] (\a)--++(\b-30:0.7); \draw[e,dotted] (\a)--++(\b:0.7); \draw[e,dotted] (\a)--++(\b+30:0.7);}
\foreach \a in {c0,c2}{\draw[e,dotted] (\a)--++(25:0.7); \draw[e,dotted] (\a)--++(0:0.7); \draw[e,dotted] (\a)--++(-25:0.7);}
\foreach \a in {d0,d1,d2}{\draw[e,dotted] (\a)--++(15:0.6); \draw[e,dotted] (\a)--++(0:0.6); \draw[e,dotted] (\a)--++(-15:0.6);}
\foreach \p in {O,b0,b1,b2,bi,c0,c1,c2,d0,d1,d2}{\node[v] at (\p){};}
\node[lab,below left=1pt] at (O){$O$};
\node[lab,below=3pt] at (b0){$B_0$};
\node[lab,right=3pt] at (b1){$B_1$};
\node[lab,above=3pt] at (b2){$B_2$};
\node[lab,right=3pt] at (bi){$B_\infty$};
\node[lab,anchor=west] at (5.55,1.1){$x$};
\node[lab,anchor=west] at (5.55,-1.1){$y$};
\node[lab,anchor=south] at (0,2.75){$w$};
\node[lab,align=left,anchor=west] at (5.9,0){$[x,y]=p^{-2}$\\[2pt] $[x,w]=1$};
\end{tikzpicture}
}
\caption{The Bruhat-Tits tree for $p=3$. The $p+1$ edges at the vertex $O$ correspond to the residue discs $B_\sigma$, $\sigma\in\mathbb P^1(\mathbb F_p)$, and the ends of the tree correspond to the points of $\mathbb P^1(\Qp)$. The chordal distance of two points is $p^{-k}$, where $k$ is the number of common edges of their paths from $O$. Here $x,y\in B_0$ have two common edges, so $[x,y]=p^{-2}$, while $w\in B_1$ and $[x,w]=1$}
\label{fig:tree}
\end{figure}

There are $p+1$ residue discs $B_\sigma$, labeled by $\sigma\in\mathbb P^1(\mathbb F_p)$, each of measure $1/(p+1)$. The disc label will be the spin of our finite model, and we define the pair weight of this model by the conditional average
\begin{equation}\label{eq:disc-average}
 W^{(1)}_a(\sigma,\sigma')
 =(p+1)^2\int_{B_\sigma}\!d\mu(x)
 \int_{B_{\sigma'}}\!d\mu(y)\,[x,y]^{-a}.
\end{equation}
It is convenient to visualize the residue discs and the chordal distance in terms of the Bruhat-Tits tree~\cite{Zabrodin:1988ep}, as shown in Fig.~\ref{fig:tree}. For different discs one has $[x,y]=1$, so this weight is equal to $1$. In the same disc the distance takes the values $p^{-k}$, $k\ge1$, with probabilities $(1-p^{-1})p^{-(k-1)}$. Hence the equal-spin weight is
\begin{equation}\label{eq:r}
 r(a)=(1-p^{-1})\sum_{k\ge1}p^{-(k-1)}p^{ka}
 =\frac{p-1}{p^{1-a}-1}.
\end{equation}
The sum converges for $\operatorname{Re}a<1$. Below we also use its meromorphic continuation. We have therefore obtained
\begin{equation}\label{eq:finite-pair}
 W^{(1)}_a(\sigma,\sigma')=1+[r(a)-1]\,\delta_{\sigma\sigma'},
 \qquad q=p+1,
\end{equation}
which is precisely the form of a Potts weight.\footnote{Note that this model is different from the $p$-adic Potts models studied by Mukhamedov, Rozikov, Khrennikov and Mendes~\cite{Mukhamedov:2004potts,Khrennikov:2007potts}, where the Hamiltonian and the Gibbs measures take values in $\Qp$. In our case the spin values are the residue discs of $\mathbb P^1(\Qp)$, while the Boltzmann weights are complex numbers.}

The Potts form of~\eqref{eq:finite-pair} is not a special property of the power function. It follows from the invariance of the chordal distance and of the measure~\eqref{eq:mu} under $PGL_2(\Zp)$.

\begin{lemma}\label{lem:potts}
Let $F(x,y)$ be a function on $\mathbb P^1(\Qp)\times\mathbb P^1(\Qp)$ which is integrable with respect to $\mu\times\mu$ and invariant under $PGL_2(\Zp)$, i.e.\ $F(k\cdot x,k\cdot y)=F(x,y)$. Then its average over the residue discs has the Potts form,
\begin{equation}\label{eq:potts-lemma}
 W_F(\sigma,\sigma')=(p+1)^2\int_{B_\sigma}\!d\mu(x)\int_{B_{\sigma'}}\!d\mu(y)\,F(x,y)=w_0+(w_1-w_0)\,\delta_{\sigma\sigma'}.
\end{equation}
\end{lemma}

\begin{proof}
The disc $B_\sigma$ is the set of points of $\mathbb P^1(\Qp)$ whose reduction modulo $p$ is equal to $\sigma\in\mathbb P^1(\mathbb F_p)$. Let $k\in PGL_2(\Zp)$ and let $\bar k\in PGL_2(\mathbb F_p)$ be its reduction modulo $p$. The reduction of $k\cdot x$ is $\bar k$ applied to the reduction of $x$, therefore $k$ maps $B_\sigma$ onto $B_{\bar k\cdot\sigma}$. Since $\mu$ and $F$ are invariant, the change of variables $x\to k\cdot x$, $y\to k\cdot y$ in~\eqref{eq:potts-lemma} gives
\begin{equation}
 W_F(\bar k\cdot\sigma,\bar k\cdot\sigma')=W_F(\sigma,\sigma').
\end{equation}
The reduction map $PGL_2(\Zp)\to PGL_2(\mathbb F_p)$ is surjective, and the group $PGL_2(\mathbb F_p)$ acts doubly transitively on $\mathbb P^1(\mathbb F_p)$. Hence $W_F(\sigma,\sigma')$ takes one value $w_0$ on all pairs with $\sigma\ne\sigma'$ and one value $w_1$ on all pairs with $\sigma=\sigma'$.
\end{proof}

In particular, any integrable function of $[x,y]$ gives a Potts weight with $q=p+1$ states. When $w_0\ne0$, dividing by $w_0$ fixes the off-diagonal weight to $1$, and the remaining parameter is $w_1/w_0$. For the weight~\eqref{eq:chordal} this ratio is the function $r(a)$. Therefore the finite star-triangle relation~\eqref{eq:W4}, which we check below, holds simply because the finite model is the Potts model, and it is not a consequence of the continuous star-triangle relation~\eqref{eq:str}. Accordingly, the invariant which we obtain is the known Potts evaluation of the Jones polynomial at $q=p+1$~\cite{Jones:1989ed,Wu:1992zzb}. The $p$-adic input consists of the number of residue discs and the function~\eqref{eq:r}, which relates the Jones variable to the exponents of the weights.

Note also that averaging individual pair weights does not by itself preserve the continuous star-triangle identity, since the weights at a star share the same integration variable. Therefore we check Wu's conditions for the finite model directly.

Next, we introduce the crossing weights
\begin{equation}\label{eq:crossing-weights}
 W_\pm(\sigma,\sigma')=c_\pm\,W^{(1)}_{a_\pm}(\sigma,\sigma').
\end{equation}
For unequal spins the pointwise inversion relation~\eqref{eq:W3} gives $c_+c_-=1$. For equal spins it then gives $r_+r_-=1$, where $r_\pm=r(a_\pm)$. The other inversion relation fixes the remaining parameter. For $\sigma\ne\sigma'$ one finds
\begin{equation}
 \sum_\rho W_-(\sigma,\rho)\,W_+(\rho,\sigma')
 =q-2+r_++r_-,
\end{equation}
since there are $q-2$ terms with $\rho$ different from both external spins, and two terms with $\rho$ equal to one of them. Thus~\eqref{eq:W2} requires $r_++r_-=2-q$. When $\sigma=\sigma'$, the sum is already equal to $q$ because $r_+r_-=1$.

\paragraph{Convolution partner} The off-diagonal condition $r_++r_-=2-q$ has a simple meaning. Indeed, from~\eqref{eq:r} one finds the identity
\begin{equation}\label{eq:r-identity}
 r(a)+r(2-a)=1-p=2-q
\end{equation}
for all $a$. Since $r$ is a Möbius function of $p^{1-a}$, the off-diagonal part of~\eqref{eq:W2} holds if and only if $a_-\equiv2-a_+$ modulo the period $\tau$. This is precisely the averaged analogue of the continuous convolution inverse. Recall that the convolution inverse of $|x|_p^{-a}$ is proportional to $|x|_p^{-(2-a)}$, see Appendix~\ref{app:continuous}. In other words, after averaging the two inversion relations become compatible. The convolution relation fixes the pair of exponents, and the pointwise relation $r(a_+)\,r(2-a_+)=1$ then fixes $a_+$.

Writing $r_+=-t_p^{-1}$ and $r_-=-t_p$, the two inversion conditions become
\begin{equation}\label{eq:tp-condition}
 t_p+t_p^{-1}=q-2=p-1.
\end{equation}
As mentioned in the Introduction, this is the usual relation for the $q$-state Potts model with $q=p+1$. In order to find the corresponding exponents, we substitute $r(a_+)=-t_p^{-1}$ and $r(a_-)=-t_p$ into~\eqref{eq:r}. Using $(p-1)t_p=t_p^2+1$, one obtains
\begin{equation}\label{eq:exponents}
 p^{1-a_\pm}=-t_p^{\pm2},\qquad
 a_\pm\equiv1\mp2\log_p t_p+\frac{\tau}{2}
 \pmod{\tau},
\end{equation}
where $\log_p t_p=\operatorname{Log}(t_p)/\ln p$ with a fixed branch of the logarithm. We choose $t_2=e^{i\pi/3}$, $t_3=1$, and the root $t_p>1$ for $p\ge5$.\footnote{The reciprocal root gives the mirror evaluation.} In agreement with~\eqref{eq:r-identity}, one has $a_++a_-\equiv2$. Therefore at least one of the diagonal weights is given by the meromorphic continuation of the sum~\eqref{eq:r}. Indeed, for $p=2,3$ both exponents have real part one, while for $p\ge5$ exactly one of them lies outside the half-plane of convergence.

The minus sign in $p^{1-a_\pm}=-t_p^{\pm2}$ has a simple and rather curious interpretation. Writing $a_\pm=b_\pm+\tau/2$ with $b_\pm=1\mp2\log_p t_p$ and using~\eqref{eq:eta}, one finds that the underlying weights are
\begin{equation}\label{eq:twisted}
 |x-y|_p^{-a_\pm}=\eta(x-y)\,|x-y|_p^{-b_\pm}.
\end{equation}
In other words, the weights which give the Jones polynomial are twisted by the unramified quadratic character. For $p\ge5$ the exponents $b_\pm$ are real and $b_++b_-=2$. For $p=3$ one has $b_\pm=1$, and for $p=2$ the exponents $b_\pm$ are complex.

\subsection{State sum for the Jones polynomial at \texorpdfstring{$t_p$}{tp}}\label{sec:jones-tp}

The remaining normalization is the Potts normalization of Appendix~\ref{app:wu}. We set $c_\pm=t_p^{\pm1/4}$. Then the crossing weights are
\begin{equation}\label{eq:crossing-values}
\renewcommand{\arraystretch}{1.4}
\begin{array}{c|cc}
 &\quad\sigma\ne\sigma'\quad&\quad\sigma=\sigma'\quad\\\hline
 W_+(\sigma,\sigma')&t_p^{1/4}&-t_p^{-3/4}\\
 W_-(\sigma,\sigma')&t_p^{-1/4}&-t_p^{3/4}
\end{array}
\end{equation}
They coincide with the Potts weights~\eqref{eq:potts} at $t=t_p$, so the finite star-triangle relation~\eqref{eq:W4} holds. Moreover, with
\begin{equation}
 \mathcal D=-(t_p^{1/2}+t_p^{-1/2}),\qquad
 \alpha=-t_p^{-3/4},\qquad \mathcal D^2=p+1,
\end{equation}
the equal-spin entries are $\alpha^{\pm1}$, and the row sums are
\begin{equation}
 \sum_{\sigma'}W_+(\sigma,\sigma')
 =t_p^{1/4}(p-t_p^{-1})=\mathcal D\alpha^{-1},\qquad
 \sum_{\sigma'}W_-(\sigma,\sigma')
 =t_p^{-1/4}(p-t_p)=\mathcal D\alpha.
\end{equation}
This verifies the curl normalization~\eqref{eq:W1}, and therefore all conditions needed for Wu's construction are satisfied.

For a knot diagram $K$, let us color its faces as in Appendix~\ref{app:wu} and assign a residue-disc label to each shaded face. Multiplying the entries of~\eqref{eq:crossing-values} over all crossings and summing over the labels, one ends up with
\begin{equation}\label{eq:padic-jones-sum}
 V_K(t_p)=\mathcal D^{-N-1}\alpha^{-w(K)}
 \sum_{\sigma_1,\ldots,\sigma_N\in\mathbb P^1(\mathbb F_p)}
 \;\prod_{e\in E(K)}W_{\epsilon_e}(\sigma_{i(e)},\sigma_{j(e)}),
\end{equation}
where $N$ is the number of shaded faces, $E(K)$ denotes the set of crossings, $i(e),j(e)$ label the two shaded faces meeting at a crossing, and $\epsilon_e$ is its checkerboard crossing type. We use the standard oriented writhe $w(K)$, with a positive braid crossing contributing $+1$. Equivalently,~\eqref{eq:padic-jones-sum} is $\mathcal D^{-1}\alpha^{-w(K)}Z$. The factor $\alpha^{-w(K)}$ removes the effect of a curl, while $\mathcal D^{-N-1}$ includes the face normalization and fixes the unknot to $1$.

Thus the $p$-adic weights supply a finite Potts model with $q=p+1$ states, and Wu's construction applied to this model gives the ordinary Jones polynomial evaluated at~\eqref{eq:tp-condition}. In agreement with Lemma~\ref{lem:potts}, the weights enter this result only through the number of states and the relation~\eqref{eq:exponents} between $t_p$ and the exponents.

\subsection{Examples: trefoil, figure-eight knot and cinquefoil}\label{sec:examples}

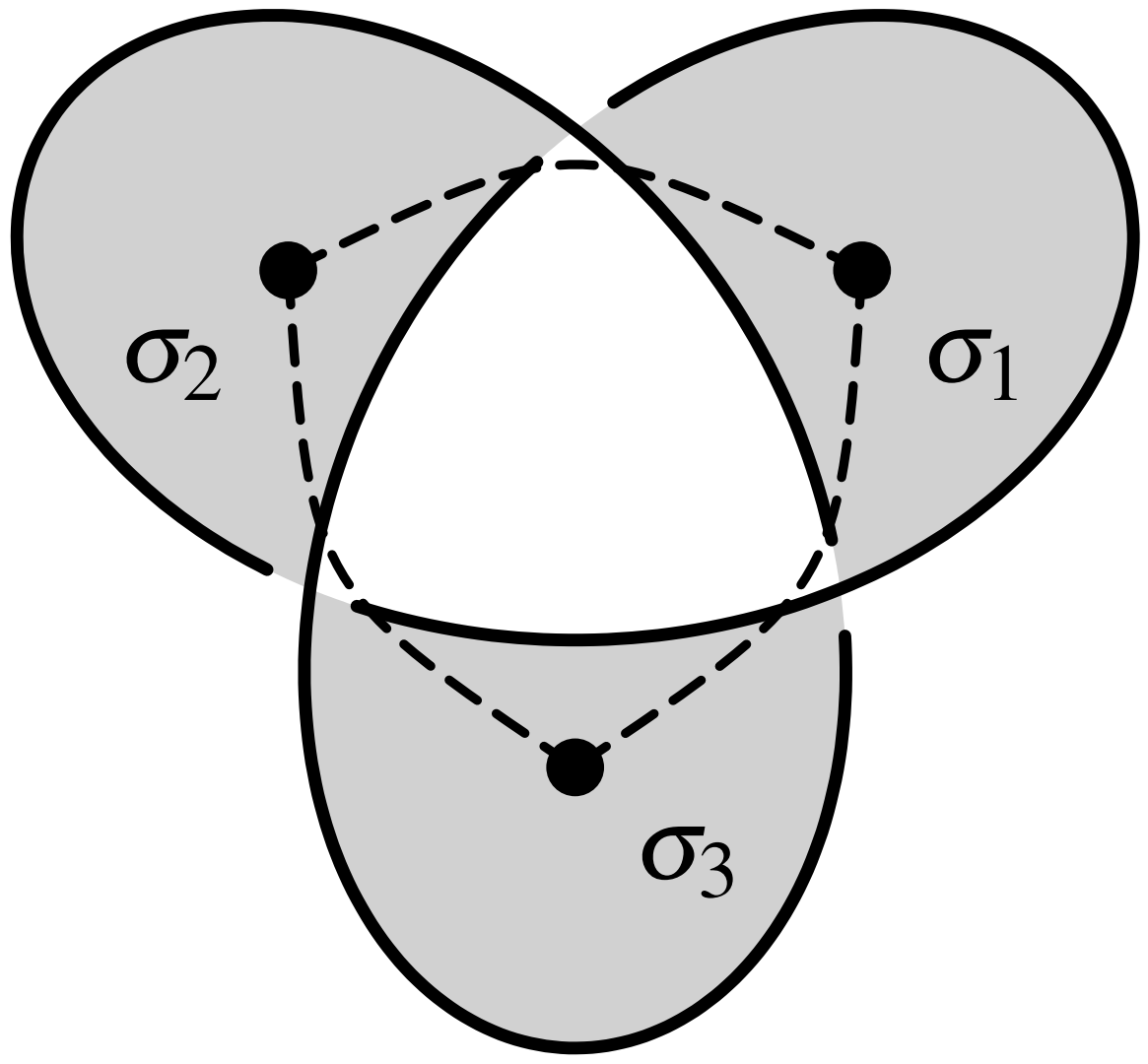
\begin{figure}[htbp]
\centering
{\sffamily
\resizebox{8.2cm}{!}{%
\setlength{\fboxsep}{4pt}\setlength{\fboxrule}{0pt}\fbox{%
\begin{tikzpicture}[scale=0.62,line join=round,line cap=round]
\fill[black!18] (1.299,-0.750)--(1.356,-0.726)--(1.412,-0.701)--(1.467,-0.675)--(1.522,-0.648)--(1.575,-0.621)--(1.628,-0.592)--(1.680,-0.563)--(1.731,-0.532)--(1.780,-0.501)--(1.829,-0.469)--(1.877,-0.436)--(1.924,-0.402)--(1.969,-0.368)--(2.014,-0.333)--(2.057,-0.297)--(2.099,-0.261)--(2.140,-0.224)--(2.180,-0.186)--(2.218,-0.148)--(2.255,-0.109)--(2.291,-0.070)--(2.326,-0.030)--(2.359,0.010)--(2.391,0.050)--(2.422,0.091)--(2.451,0.133)--(2.479,0.174)--(2.506,0.216)--(2.531,0.259)--(2.555,0.301)--(2.577,0.344)--(2.598,0.387)--(2.617,0.430)--(2.635,0.473)--(2.652,0.516)--(2.667,0.559)--(2.680,0.602)--(2.692,0.645)--(2.703,0.688)--(2.712,0.731)--(2.720,0.774)--(2.726,0.817)--(2.731,0.859)--(2.734,0.901)--(2.735,0.943)--(2.736,0.985)--(2.734,1.026)--(2.732,1.067)--(2.728,1.108)--(2.722,1.148)--(2.715,1.187)--(2.706,1.226)--(2.697,1.265)--(2.685,1.303)--(2.672,1.340)--(2.658,1.377)--(2.643,1.413)--(2.626,1.448)--(2.608,1.483)--(2.588,1.517)--(2.567,1.550)--(2.545,1.582)--(2.522,1.614)--(2.497,1.644)--(2.471,1.674)--(2.444,1.703)--(2.415,1.731)--(2.386,1.758)--(2.355,1.784)--(2.323,1.808)--(2.290,1.832)--(2.256,1.855)--(2.221,1.877)--(2.185,1.897)--(2.148,1.917)--(2.109,1.935)--(2.070,1.952)--(2.030,1.968)--(1.989,1.983)--(1.948,1.997)--(1.905,2.009)--(1.862,2.020)--(1.818,2.030)--(1.773,2.038)--(1.727,2.046)--(1.681,2.052)--(1.634,2.056)--(1.586,2.060)--(1.538,2.062)--(1.490,2.062)--(1.440,2.062)--(1.391,2.060)--(1.341,2.057)--(1.290,2.052)--(1.239,2.046)--(1.188,2.038)--(1.137,2.030)--(1.085,2.019)--(1.033,2.008)--(0.981,1.995)--(0.928,1.981)--(0.876,1.965)--(0.824,1.948)--(0.771,1.930)--(0.718,1.910)--(0.666,1.889)--(0.613,1.867)--(0.561,1.844)--(0.509,1.819)--(0.456,1.792)--(0.404,1.765)--(0.353,1.736)--(0.301,1.706)--(0.250,1.675)--(0.199,1.642)--(0.149,1.608)--(0.099,1.573)--(0.049,1.537)--(0.000,1.500)--(0.000,1.500)--(0.017,1.487)--(0.034,1.473)--(0.051,1.460)--(0.068,1.446)--(0.085,1.432)--(0.101,1.418)--(0.118,1.404)--(0.135,1.390)--(0.151,1.375)--(0.168,1.361)--(0.184,1.346)--(0.200,1.331)--(0.216,1.317)--(0.233,1.302)--(0.249,1.286)--(0.265,1.271)--(0.280,1.256)--(0.296,1.240)--(0.312,1.224)--(0.328,1.209)--(0.343,1.193)--(0.359,1.177)--(0.374,1.161)--(0.389,1.144)--(0.405,1.128)--(0.420,1.111)--(0.435,1.095)--(0.450,1.078)--(0.465,1.061)--(0.479,1.044)--(0.494,1.027)--(0.509,1.010)--(0.523,0.993)--(0.537,0.975)--(0.552,0.958)--(0.566,0.940)--(0.580,0.923)--(0.594,0.905)--(0.607,0.887)--(0.621,0.869)--(0.635,0.851)--(0.648,0.833)--(0.662,0.814)--(0.675,0.796)--(0.688,0.778)--(0.701,0.759)--(0.714,0.740)--(0.727,0.722)--(0.740,0.703)--(0.752,0.684)--(0.765,0.665)--(0.777,0.646)--(0.789,0.627)--(0.802,0.607)--(0.814,0.588)--(0.825,0.569)--(0.837,0.549)--(0.849,0.529)--(0.860,0.510)--(0.872,0.490)--(0.883,0.470)--(0.894,0.450)--(0.905,0.431)--(0.916,0.411)--(0.927,0.391)--(0.937,0.370)--(0.948,0.350)--(0.958,0.330)--(0.968,0.310)--(0.978,0.289)--(0.988,0.269)--(0.998,0.248)--(1.008,0.228)--(1.017,0.207)--(1.027,0.187)--(1.036,0.166)--(1.045,0.145)--(1.054,0.124)--(1.063,0.103)--(1.072,0.083)--(1.080,0.062)--(1.089,0.041)--(1.097,0.020)--(1.105,-0.001)--(1.113,-0.022)--(1.121,-0.044)--(1.129,-0.065)--(1.137,-0.086)--(1.144,-0.107)--(1.151,-0.128)--(1.159,-0.150)--(1.166,-0.171)--(1.172,-0.192)--(1.179,-0.214)--(1.186,-0.235)--(1.192,-0.256)--(1.198,-0.278)--(1.205,-0.299)--(1.211,-0.321)--(1.216,-0.342)--(1.222,-0.363)--(1.228,-0.385)--(1.233,-0.406)--(1.238,-0.428)--(1.243,-0.449)--(1.248,-0.471)--(1.253,-0.492)--(1.258,-0.514)--(1.262,-0.535)--(1.267,-0.557)--(1.271,-0.578)--(1.275,-0.600)--(1.279,-0.621)--(1.283,-0.643)--(1.286,-0.664)--(1.290,-0.686)--(1.293,-0.707)--(1.296,-0.729)--(1.299,-0.750)--cycle;
\fill[black!18] (0.000,1.500)--(-0.017,1.487)--(-0.034,1.473)--(-0.051,1.460)--(-0.068,1.446)--(-0.085,1.432)--(-0.101,1.418)--(-0.118,1.404)--(-0.135,1.390)--(-0.151,1.375)--(-0.168,1.361)--(-0.184,1.346)--(-0.200,1.331)--(-0.216,1.317)--(-0.233,1.302)--(-0.249,1.286)--(-0.265,1.271)--(-0.280,1.256)--(-0.296,1.240)--(-0.312,1.224)--(-0.328,1.209)--(-0.343,1.193)--(-0.359,1.177)--(-0.374,1.161)--(-0.389,1.144)--(-0.405,1.128)--(-0.420,1.111)--(-0.435,1.095)--(-0.450,1.078)--(-0.465,1.061)--(-0.479,1.044)--(-0.494,1.027)--(-0.509,1.010)--(-0.523,0.993)--(-0.537,0.975)--(-0.552,0.958)--(-0.566,0.940)--(-0.580,0.923)--(-0.594,0.905)--(-0.607,0.887)--(-0.621,0.869)--(-0.635,0.851)--(-0.648,0.833)--(-0.662,0.814)--(-0.675,0.796)--(-0.688,0.778)--(-0.701,0.759)--(-0.714,0.740)--(-0.727,0.722)--(-0.740,0.703)--(-0.752,0.684)--(-0.765,0.665)--(-0.777,0.646)--(-0.789,0.627)--(-0.802,0.607)--(-0.814,0.588)--(-0.825,0.569)--(-0.837,0.549)--(-0.849,0.529)--(-0.860,0.510)--(-0.872,0.490)--(-0.883,0.470)--(-0.894,0.450)--(-0.905,0.431)--(-0.916,0.411)--(-0.927,0.391)--(-0.937,0.370)--(-0.948,0.350)--(-0.958,0.330)--(-0.968,0.310)--(-0.978,0.289)--(-0.988,0.269)--(-0.998,0.248)--(-1.008,0.228)--(-1.017,0.207)--(-1.027,0.187)--(-1.036,0.166)--(-1.045,0.145)--(-1.054,0.124)--(-1.063,0.103)--(-1.072,0.083)--(-1.080,0.062)--(-1.089,0.041)--(-1.097,0.020)--(-1.105,-0.001)--(-1.113,-0.022)--(-1.121,-0.044)--(-1.129,-0.065)--(-1.137,-0.086)--(-1.144,-0.107)--(-1.151,-0.128)--(-1.159,-0.150)--(-1.166,-0.171)--(-1.172,-0.192)--(-1.179,-0.214)--(-1.186,-0.235)--(-1.192,-0.256)--(-1.198,-0.278)--(-1.205,-0.299)--(-1.211,-0.321)--(-1.216,-0.342)--(-1.222,-0.363)--(-1.228,-0.385)--(-1.233,-0.406)--(-1.238,-0.428)--(-1.243,-0.449)--(-1.248,-0.471)--(-1.253,-0.492)--(-1.258,-0.514)--(-1.262,-0.535)--(-1.267,-0.557)--(-1.271,-0.578)--(-1.275,-0.600)--(-1.279,-0.621)--(-1.283,-0.643)--(-1.286,-0.664)--(-1.290,-0.686)--(-1.293,-0.707)--(-1.296,-0.729)--(-1.299,-0.750)--(-1.299,-0.750)--(-1.356,-0.726)--(-1.412,-0.701)--(-1.467,-0.675)--(-1.522,-0.648)--(-1.575,-0.621)--(-1.628,-0.592)--(-1.680,-0.563)--(-1.731,-0.532)--(-1.780,-0.501)--(-1.829,-0.469)--(-1.877,-0.436)--(-1.924,-0.402)--(-1.969,-0.368)--(-2.014,-0.333)--(-2.057,-0.297)--(-2.099,-0.261)--(-2.140,-0.224)--(-2.180,-0.186)--(-2.218,-0.148)--(-2.255,-0.109)--(-2.291,-0.070)--(-2.326,-0.030)--(-2.359,0.010)--(-2.391,0.050)--(-2.422,0.091)--(-2.451,0.133)--(-2.479,0.174)--(-2.506,0.216)--(-2.531,0.259)--(-2.555,0.301)--(-2.577,0.344)--(-2.598,0.387)--(-2.617,0.430)--(-2.635,0.473)--(-2.652,0.516)--(-2.667,0.559)--(-2.680,0.602)--(-2.692,0.645)--(-2.703,0.688)--(-2.712,0.731)--(-2.720,0.774)--(-2.726,0.817)--(-2.731,0.859)--(-2.734,0.901)--(-2.735,0.943)--(-2.736,0.985)--(-2.734,1.026)--(-2.732,1.067)--(-2.728,1.108)--(-2.722,1.148)--(-2.715,1.187)--(-2.706,1.226)--(-2.697,1.265)--(-2.685,1.303)--(-2.672,1.340)--(-2.658,1.377)--(-2.643,1.413)--(-2.626,1.448)--(-2.608,1.483)--(-2.588,1.517)--(-2.567,1.550)--(-2.545,1.582)--(-2.522,1.614)--(-2.497,1.644)--(-2.471,1.674)--(-2.444,1.703)--(-2.415,1.731)--(-2.386,1.758)--(-2.355,1.784)--(-2.323,1.808)--(-2.290,1.832)--(-2.256,1.855)--(-2.221,1.877)--(-2.185,1.897)--(-2.148,1.917)--(-2.109,1.935)--(-2.070,1.952)--(-2.030,1.968)--(-1.989,1.983)--(-1.948,1.997)--(-1.905,2.009)--(-1.862,2.020)--(-1.818,2.030)--(-1.773,2.038)--(-1.727,2.046)--(-1.681,2.052)--(-1.634,2.056)--(-1.586,2.060)--(-1.538,2.062)--(-1.490,2.062)--(-1.440,2.062)--(-1.391,2.060)--(-1.341,2.057)--(-1.290,2.052)--(-1.239,2.046)--(-1.188,2.038)--(-1.137,2.030)--(-1.085,2.019)--(-1.033,2.008)--(-0.981,1.995)--(-0.928,1.981)--(-0.876,1.965)--(-0.824,1.948)--(-0.771,1.930)--(-0.718,1.910)--(-0.666,1.889)--(-0.613,1.867)--(-0.561,1.844)--(-0.509,1.819)--(-0.456,1.792)--(-0.404,1.765)--(-0.353,1.736)--(-0.301,1.706)--(-0.250,1.675)--(-0.199,1.642)--(-0.149,1.608)--(-0.099,1.573)--(-0.049,1.537)--(0.000,1.500)--cycle;
\fill[black!18] (-1.299,-0.750)--(-1.307,-0.811)--(-1.313,-0.872)--(-1.319,-0.933)--(-1.323,-0.994)--(-1.325,-1.054)--(-1.327,-1.114)--(-1.327,-1.174)--(-1.326,-1.233)--(-1.324,-1.291)--(-1.321,-1.350)--(-1.316,-1.407)--(-1.310,-1.465)--(-1.303,-1.521)--(-1.295,-1.577)--(-1.286,-1.633)--(-1.276,-1.687)--(-1.264,-1.741)--(-1.251,-1.795)--(-1.237,-1.847)--(-1.222,-1.899)--(-1.206,-1.949)--(-1.189,-1.999)--(-1.171,-2.048)--(-1.152,-2.096)--(-1.132,-2.143)--(-1.111,-2.189)--(-1.089,-2.234)--(-1.065,-2.278)--(-1.041,-2.321)--(-1.016,-2.363)--(-0.991,-2.404)--(-0.964,-2.443)--(-0.936,-2.481)--(-0.908,-2.519)--(-0.879,-2.554)--(-0.849,-2.589)--(-0.819,-2.622)--(-0.787,-2.654)--(-0.755,-2.685)--(-0.723,-2.714)--(-0.689,-2.742)--(-0.656,-2.769)--(-0.621,-2.794)--(-0.586,-2.818)--(-0.551,-2.841)--(-0.515,-2.862)--(-0.478,-2.881)--(-0.442,-2.899)--(-0.405,-2.916)--(-0.367,-2.931)--(-0.329,-2.945)--(-0.291,-2.957)--(-0.253,-2.968)--(-0.214,-2.977)--(-0.176,-2.985)--(-0.137,-2.991)--(-0.098,-2.995)--(-0.059,-2.998)--(-0.020,-3.000)--(0.020,-3.000)--(0.059,-2.998)--(0.098,-2.995)--(0.137,-2.991)--(0.176,-2.985)--(0.214,-2.977)--(0.253,-2.968)--(0.291,-2.957)--(0.329,-2.945)--(0.367,-2.931)--(0.405,-2.916)--(0.442,-2.899)--(0.478,-2.881)--(0.515,-2.862)--(0.551,-2.841)--(0.586,-2.818)--(0.621,-2.794)--(0.656,-2.769)--(0.689,-2.742)--(0.723,-2.714)--(0.755,-2.685)--(0.787,-2.654)--(0.819,-2.622)--(0.849,-2.589)--(0.879,-2.554)--(0.908,-2.519)--(0.936,-2.481)--(0.964,-2.443)--(0.991,-2.404)--(1.016,-2.363)--(1.041,-2.321)--(1.065,-2.278)--(1.089,-2.234)--(1.111,-2.189)--(1.132,-2.143)--(1.152,-2.096)--(1.171,-2.048)--(1.189,-1.999)--(1.206,-1.949)--(1.222,-1.899)--(1.237,-1.847)--(1.251,-1.795)--(1.264,-1.741)--(1.276,-1.687)--(1.286,-1.633)--(1.295,-1.577)--(1.303,-1.521)--(1.310,-1.465)--(1.316,-1.407)--(1.321,-1.350)--(1.324,-1.291)--(1.326,-1.233)--(1.327,-1.174)--(1.327,-1.114)--(1.325,-1.054)--(1.323,-0.994)--(1.319,-0.933)--(1.313,-0.872)--(1.307,-0.811)--(1.299,-0.750)--(1.299,-0.750)--(1.279,-0.758)--(1.259,-0.766)--(1.239,-0.774)--(1.218,-0.782)--(1.198,-0.789)--(1.177,-0.797)--(1.157,-0.804)--(1.136,-0.811)--(1.116,-0.819)--(1.095,-0.826)--(1.074,-0.832)--(1.053,-0.839)--(1.032,-0.846)--(1.011,-0.852)--(0.990,-0.858)--(0.968,-0.865)--(0.947,-0.871)--(0.926,-0.877)--(0.904,-0.882)--(0.883,-0.888)--(0.861,-0.894)--(0.840,-0.899)--(0.818,-0.904)--(0.796,-0.909)--(0.775,-0.914)--(0.753,-0.919)--(0.731,-0.924)--(0.709,-0.929)--(0.687,-0.933)--(0.665,-0.937)--(0.643,-0.941)--(0.621,-0.945)--(0.598,-0.949)--(0.576,-0.953)--(0.554,-0.957)--(0.532,-0.960)--(0.509,-0.963)--(0.487,-0.967)--(0.464,-0.970)--(0.442,-0.972)--(0.419,-0.975)--(0.397,-0.978)--(0.374,-0.980)--(0.352,-0.983)--(0.329,-0.985)--(0.307,-0.987)--(0.284,-0.989)--(0.261,-0.990)--(0.239,-0.992)--(0.216,-0.993)--(0.193,-0.995)--(0.171,-0.996)--(0.148,-0.997)--(0.125,-0.998)--(0.102,-0.999)--(0.080,-0.999)--(0.057,-1.000)--(0.034,-1.000)--(0.011,-1.000)--(-0.011,-1.000)--(-0.034,-1.000)--(-0.057,-1.000)--(-0.080,-0.999)--(-0.102,-0.999)--(-0.125,-0.998)--(-0.148,-0.997)--(-0.171,-0.996)--(-0.193,-0.995)--(-0.216,-0.993)--(-0.239,-0.992)--(-0.261,-0.990)--(-0.284,-0.989)--(-0.307,-0.987)--(-0.329,-0.985)--(-0.352,-0.983)--(-0.374,-0.980)--(-0.397,-0.978)--(-0.419,-0.975)--(-0.442,-0.972)--(-0.464,-0.970)--(-0.487,-0.967)--(-0.509,-0.963)--(-0.532,-0.960)--(-0.554,-0.957)--(-0.576,-0.953)--(-0.598,-0.949)--(-0.621,-0.945)--(-0.643,-0.941)--(-0.665,-0.937)--(-0.687,-0.933)--(-0.709,-0.929)--(-0.731,-0.924)--(-0.753,-0.919)--(-0.775,-0.914)--(-0.796,-0.909)--(-0.818,-0.904)--(-0.840,-0.899)--(-0.861,-0.894)--(-0.883,-0.888)--(-0.904,-0.882)--(-0.926,-0.877)--(-0.947,-0.871)--(-0.968,-0.865)--(-0.990,-0.858)--(-1.011,-0.852)--(-1.032,-0.846)--(-1.053,-0.839)--(-1.074,-0.832)--(-1.095,-0.826)--(-1.116,-0.819)--(-1.136,-0.811)--(-1.157,-0.804)--(-1.177,-0.797)--(-1.198,-0.789)--(-1.218,-0.782)--(-1.239,-0.774)--(-1.259,-0.766)--(-1.279,-0.758)--(-1.299,-0.750)--cycle;
\draw[line width=1.1pt] (-1.070,-0.834) -- (-1.012,-0.852) -- (-0.953,-0.869) -- (-0.894,-0.885) -- (-0.834,-0.900) -- (-0.774,-0.914) -- (-0.714,-0.927) -- (-0.653,-0.939) -- (-0.592,-0.950) -- (-0.531,-0.960) -- (-0.469,-0.969) -- (-0.407,-0.977) -- (-0.345,-0.983) -- (-0.282,-0.989) -- (-0.220,-0.993) -- (-0.157,-0.997) -- (-0.094,-0.999) -- (-0.031,-1.000) -- (0.031,-1.000) -- (0.094,-0.999) -- (0.157,-0.997) -- (0.220,-0.993) -- (0.282,-0.989) -- (0.345,-0.983) -- (0.407,-0.977) -- (0.469,-0.969) -- (0.531,-0.960) -- (0.592,-0.950) -- (0.653,-0.939) -- (0.714,-0.927) -- (0.774,-0.914) -- (0.834,-0.900) -- (0.894,-0.885) -- (0.953,-0.869) -- (1.012,-0.852) -- (1.070,-0.834) -- (1.127,-0.815) -- (1.184,-0.794) -- (1.240,-0.773) -- (1.296,-0.751) -- (1.351,-0.728) -- (1.405,-0.704) -- (1.458,-0.680) -- (1.511,-0.654) -- (1.563,-0.628) -- (1.614,-0.600) -- (1.664,-0.572) -- (1.713,-0.543) -- (1.761,-0.513) -- (1.809,-0.483) -- (1.855,-0.451) -- (1.901,-0.419) -- (1.945,-0.387) -- (1.988,-0.353) -- (2.031,-0.319) -- (2.072,-0.285) -- (2.112,-0.249) -- (2.151,-0.213) -- (2.189,-0.177) -- (2.226,-0.140) -- (2.262,-0.103) -- (2.296,-0.065) -- (2.329,-0.027) -- (2.361,0.012) -- (2.392,0.051) -- (2.422,0.091) -- (2.450,0.131) -- (2.477,0.171) -- (2.503,0.211) -- (2.527,0.252) -- (2.550,0.293) -- (2.572,0.334) -- (2.592,0.375) -- (2.611,0.416) -- (2.629,0.458) -- (2.646,0.499) -- (2.661,0.541) -- (2.674,0.583) -- (2.687,0.624) -- (2.698,0.666) -- (2.707,0.707) -- (2.715,0.748) -- (2.722,0.790) -- (2.728,0.831) -- (2.732,0.871) -- (2.734,0.912) -- (2.736,0.952) -- (2.736,0.992) -- (2.734,1.032) -- (2.731,1.071) -- (2.727,1.110) -- (2.722,1.149) -- (2.715,1.187) -- (2.707,1.225) -- (2.697,1.262) -- (2.687,1.298) -- (2.675,1.334) -- (2.661,1.370) -- (2.646,1.405) -- (2.631,1.439) -- (2.613,1.473) -- (2.595,1.505) -- (2.575,1.538) -- (2.554,1.569) -- (2.532,1.600) -- (2.509,1.630) -- (2.485,1.659) -- (2.459,1.687) -- (2.432,1.714) -- (2.405,1.741) -- (2.376,1.766) -- (2.346,1.791) -- (2.315,1.815) -- (2.283,1.837) -- (2.250,1.859) -- (2.216,1.880) -- (2.181,1.899) -- (2.145,1.918) -- (2.108,1.936) -- (2.070,1.952) -- (2.032,1.968) -- (1.993,1.982) -- (1.952,1.995) -- (1.912,2.007) -- (1.870,2.018) -- (1.827,2.028) -- (1.784,2.036) -- (1.741,2.044) -- (1.696,2.050) -- (1.651,2.055) -- (1.606,2.059) -- (1.560,2.061) -- (1.513,2.062) -- (1.466,2.062) -- (1.418,2.061) -- (1.370,2.059) -- (1.322,2.055) -- (1.273,2.050) -- (1.224,2.044) -- (1.175,2.036) -- (1.125,2.027) -- (1.075,2.017) -- (1.025,2.006) -- (0.975,1.994) -- (0.924,1.980) -- (0.874,1.965) -- (0.823,1.948) -- (0.773,1.931) -- (0.722,1.912) -- (0.671,1.892) -- (0.621,1.870) -- (0.570,1.848) -- (0.520,1.824) -- (0.470,1.799) -- (0.420,1.773) -- (0.370,1.746) -- (0.320,1.717) -- (0.271,1.687) -- (0.222,1.657) -- (0.188,1.635);
\draw[line width=1.1pt] (-0.187,1.343) -- (-0.232,1.302) -- (-0.276,1.260) -- (-0.320,1.217) -- (-0.362,1.173) -- (-0.405,1.128) -- (-0.446,1.082) -- (-0.487,1.035) -- (-0.527,0.988) -- (-0.566,0.940) -- (-0.605,0.891) -- (-0.642,0.841) -- (-0.679,0.790) -- (-0.715,0.739) -- (-0.750,0.687) -- (-0.785,0.634) -- (-0.818,0.581) -- (-0.850,0.527) -- (-0.882,0.473) -- (-0.912,0.418) -- (-0.942,0.362) -- (-0.970,0.306) -- (-0.997,0.250) -- (-1.024,0.193) -- (-1.049,0.136) -- (-1.074,0.078) -- (-1.097,0.021) -- (-1.119,-0.038) -- (-1.140,-0.096) -- (-1.160,-0.155) -- (-1.179,-0.213) -- (-1.197,-0.273) -- (-1.214,-0.332) -- (-1.229,-0.391) -- (-1.244,-0.450) -- (-1.257,-0.510) -- (-1.269,-0.569) -- (-1.280,-0.628) -- (-1.290,-0.687) -- (-1.299,-0.747) -- (-1.306,-0.806) -- (-1.312,-0.864) -- (-1.318,-0.923) -- (-1.322,-0.981) -- (-1.325,-1.040) -- (-1.327,-1.097) -- (-1.327,-1.155) -- (-1.327,-1.212) -- (-1.325,-1.269) -- (-1.322,-1.325) -- (-1.318,-1.381) -- (-1.313,-1.436) -- (-1.307,-1.491) -- (-1.300,-1.545) -- (-1.292,-1.599) -- (-1.282,-1.652) -- (-1.272,-1.705) -- (-1.261,-1.756) -- (-1.248,-1.807) -- (-1.234,-1.858) -- (-1.220,-1.907) -- (-1.204,-1.956) -- (-1.188,-2.004) -- (-1.170,-2.051) -- (-1.152,-2.097) -- (-1.132,-2.143) -- (-1.112,-2.187) -- (-1.091,-2.230) -- (-1.068,-2.273) -- (-1.045,-2.314) -- (-1.021,-2.355) -- (-0.997,-2.394) -- (-0.971,-2.433) -- (-0.945,-2.470) -- (-0.918,-2.506) -- (-0.890,-2.541) -- (-0.862,-2.575) -- (-0.833,-2.607) -- (-0.803,-2.639) -- (-0.772,-2.669) -- (-0.741,-2.698) -- (-0.709,-2.726) -- (-0.677,-2.752) -- (-0.644,-2.777) -- (-0.611,-2.801) -- (-0.577,-2.824) -- (-0.543,-2.845) -- (-0.508,-2.865) -- (-0.473,-2.884) -- (-0.438,-2.901) -- (-0.402,-2.917) -- (-0.366,-2.932) -- (-0.330,-2.945) -- (-0.293,-2.957) -- (-0.256,-2.967) -- (-0.219,-2.976) -- (-0.182,-2.983) -- (-0.144,-2.990) -- (-0.107,-2.994) -- (-0.069,-2.998) -- (-0.031,-3.000) -- (0.006,-3.000) -- (0.044,-2.999) -- (0.082,-2.997) -- (0.119,-2.993) -- (0.157,-2.988) -- (0.194,-2.981) -- (0.231,-2.973) -- (0.268,-2.964) -- (0.305,-2.953) -- (0.342,-2.941) -- (0.378,-2.927) -- (0.414,-2.912) -- (0.450,-2.896) -- (0.485,-2.878) -- (0.520,-2.859) -- (0.555,-2.838) -- (0.589,-2.817) -- (0.622,-2.794) -- (0.655,-2.769) -- (0.688,-2.744) -- (0.720,-2.717) -- (0.752,-2.688) -- (0.783,-2.659) -- (0.813,-2.628) -- (0.842,-2.597) -- (0.871,-2.564) -- (0.900,-2.529) -- (0.927,-2.494) -- (0.954,-2.457) -- (0.980,-2.420) -- (1.005,-2.381) -- (1.030,-2.341) -- (1.053,-2.301) -- (1.076,-2.259) -- (1.098,-2.216) -- (1.119,-2.172) -- (1.139,-2.128) -- (1.158,-2.082) -- (1.176,-2.035) -- (1.193,-1.988) -- (1.210,-1.940) -- (1.225,-1.891) -- (1.239,-1.841) -- (1.252,-1.790) -- (1.264,-1.739) -- (1.276,-1.687) -- (1.286,-1.635) -- (1.295,-1.581) -- (1.303,-1.527) -- (1.309,-1.473) -- (1.315,-1.418) -- (1.320,-1.362) -- (1.323,-1.306) -- (1.326,-1.250) -- (1.327,-1.193) -- (1.327,-1.136) -- (1.326,-1.078) -- (1.324,-1.020) -- (1.322,-0.980);
\draw[line width=1.1pt] (1.257,-0.510) -- (1.244,-0.450) -- (1.229,-0.391) -- (1.214,-0.332) -- (1.197,-0.273) -- (1.179,-0.213) -- (1.160,-0.155) -- (1.140,-0.096) -- (1.119,-0.038) -- (1.097,0.021) -- (1.074,0.078) -- (1.049,0.136) -- (1.024,0.193) -- (0.997,0.250) -- (0.970,0.306) -- (0.942,0.362) -- (0.912,0.418) -- (0.882,0.473) -- (0.850,0.527) -- (0.818,0.581) -- (0.785,0.634) -- (0.750,0.687) -- (0.715,0.739) -- (0.679,0.790) -- (0.642,0.841) -- (0.605,0.891) -- (0.566,0.940) -- (0.527,0.988) -- (0.487,1.035) -- (0.446,1.082) -- (0.405,1.128) -- (0.362,1.173) -- (0.320,1.217) -- (0.276,1.260) -- (0.232,1.302) -- (0.187,1.343) -- (0.142,1.383) -- (0.096,1.423) -- (0.050,1.461) -- (0.003,1.498) -- (-0.045,1.534) -- (-0.092,1.569) -- (-0.140,1.603) -- (-0.189,1.635) -- (-0.238,1.667) -- (-0.287,1.697) -- (-0.337,1.727) -- (-0.386,1.755) -- (-0.436,1.782) -- (-0.486,1.808) -- (-0.537,1.832) -- (-0.587,1.856) -- (-0.638,1.878) -- (-0.688,1.899) -- (-0.739,1.918) -- (-0.790,1.937) -- (-0.840,1.954) -- (-0.891,1.970) -- (-0.941,1.984) -- (-0.992,1.998) -- (-1.042,2.010) -- (-1.092,2.021) -- (-1.142,2.031) -- (-1.191,2.039) -- (-1.241,2.046) -- (-1.289,2.052) -- (-1.338,2.056) -- (-1.386,2.060) -- (-1.434,2.062) -- (-1.482,2.062) -- (-1.529,2.062) -- (-1.575,2.060) -- (-1.621,2.057) -- (-1.666,2.053) -- (-1.711,2.048) -- (-1.755,2.041) -- (-1.799,2.034) -- (-1.842,2.025) -- (-1.884,2.015) -- (-1.925,2.003) -- (-1.966,1.991) -- (-2.006,1.977) -- (-2.045,1.963) -- (-2.083,1.947) -- (-2.120,1.930) -- (-2.157,1.912) -- (-2.193,1.893) -- (-2.227,1.873) -- (-2.261,1.852) -- (-2.294,1.830) -- (-2.325,1.807) -- (-2.356,1.783) -- (-2.385,1.758) -- (-2.414,1.732) -- (-2.441,1.705) -- (-2.468,1.678) -- (-2.493,1.649) -- (-2.517,1.620) -- (-2.540,1.590) -- (-2.561,1.559) -- (-2.582,1.527) -- (-2.601,1.495) -- (-2.619,1.461) -- (-2.636,1.428) -- (-2.652,1.393) -- (-2.666,1.358) -- (-2.679,1.322) -- (-2.690,1.286) -- (-2.701,1.249) -- (-2.710,1.212) -- (-2.717,1.174) -- (-2.724,1.136) -- (-2.729,1.097) -- (-2.733,1.058) -- (-2.735,1.019) -- (-2.736,0.979) -- (-2.735,0.939) -- (-2.734,0.898) -- (-2.730,0.858) -- (-2.726,0.817) -- (-2.720,0.776) -- (-2.713,0.735) -- (-2.704,0.693) -- (-2.694,0.652) -- (-2.683,0.610) -- (-2.670,0.569) -- (-2.656,0.527) -- (-2.640,0.486) -- (-2.623,0.444) -- (-2.605,0.403) -- (-2.586,0.361) -- (-2.565,0.320) -- (-2.543,0.279) -- (-2.519,0.238) -- (-2.494,0.198) -- (-2.468,0.157) -- (-2.441,0.117) -- (-2.412,0.078) -- (-2.382,0.038) -- (-2.351,-0.001) -- (-2.318,-0.039) -- (-2.285,-0.078) -- (-2.250,-0.115) -- (-2.214,-0.153) -- (-2.177,-0.189) -- (-2.138,-0.225) -- (-2.099,-0.261) -- (-2.058,-0.296) -- (-2.017,-0.331) -- (-1.974,-0.364) -- (-1.930,-0.398) -- (-1.885,-0.430) -- (-1.840,-0.462) -- (-1.793,-0.493) -- (-1.745,-0.523) -- (-1.697,-0.553) -- (-1.647,-0.581) -- (-1.597,-0.609) -- (-1.545,-0.636) -- (-1.510,-0.655);
\draw[line width=.8pt,dash pattern=on 3pt off 2pt] (1.406,0.812) .. controls (0.005,1.504) .. (-1.406,0.812);
\draw[line width=.8pt,dash pattern=on 3pt off 2pt] (1.406,0.812) .. controls (1.296,-0.751) .. (0.000,-1.624);
\draw[line width=.8pt,dash pattern=on 3pt off 2pt] (-1.406,0.812) .. controls (-1.299,-0.747) .. (0.000,-1.624);
\node[circle,fill=black,inner sep=0pt,minimum size=5pt,label={[font=\small]below right:$\sigma_1$}] at (1.406,0.812){};
\node[circle,fill=black,inner sep=0pt,minimum size=5pt,label={[font=\small]below left:$\sigma_2$}] at (-1.406,0.812){};
\node[circle,fill=black,inner sep=0pt,minimum size=5pt,label={[font=\small]below right:$\sigma_3$}] at (0.000,-1.624){};
\end{tikzpicture}}}
}
\caption{Checkerboard coloring of the trefoil diagram of Section~\ref{sec:examples}. The three shaded faces carry the spins $\sigma_1,\sigma_2,\sigma_3$, and each crossing gives an edge (dashed line) of the shaded graph, which is a triangle}
\label{fig:trefoil}
\end{figure}

Here we present a few examples of the state sum~\eqref{eq:padic-jones-sum}. As a first example, we consider the trefoil diagram whose shaded graph is a triangle with three $W_+$ edges, see Fig.~\ref{fig:trefoil}. With the standard oriented writhe this right-handed trefoil has $w=+3$. With $v=-(1+t^{-1})$ one finds
\begin{equation}
 \sum_{\text{spins}}\prod W_+=t^{3/4}
 \left(q^3+3vq^2+3v^2q+v^3q\right).
\end{equation}
Since $\mathcal D^{-1}Z=\mathcal D^{-4}\sum\prod W_+$, equation~\eqref{eq:jones} gives
\begin{equation}
 V_{3_1}(t)=t+t^3-t^4.
\end{equation}
The mirror polynomial is obtained by $t\mapsto t^{-1}$. For the figure-eight knot and the cinquefoil, respectively,
\begin{align}
 V_{4_1}(t)&=t^{-2}-t^{-1}+1-t+t^2,\\
 V_{5_1}(t)&=t^2+t^4-t^5+t^6-t^7.
\end{align}
In particular,
\begin{equation}
 V_{4_1}(t_p)=(p-1)^2-(p-1)-1=p^2-3p+1.
\end{equation}
The values of these polynomials at $t_p$ for the first few primes are collected in Table~\ref{tab:values}. For $p=3$ one has $t_p=1$, and hence $V_K(1)=1$ for any knot. Further examples could be obtained from the repeating chain knots studied by Wu and Wang~\cite{Wu:2001zeros}, by evaluating their Jones polynomials at $t_p$.

\begin{table}[t]
\caption{Values of the Jones polynomial at $t_p$, $t_p+t_p^{-1}=p-1$, for the trefoil, the figure-eight knot and the cinquefoil}
\label{tab:values}
\centering
\renewcommand{\arraystretch}{1.2}
\begin{tabular}{c|c|rrr}
$p$ & $t_p$ & $V_{3_1}(t_p)$ & $V_{4_1}(t_p)$ & $V_{5_1}(t_p)$\\\hline
2 & $e^{i\pi/3}$ & $i\sqrt3$ & $-1$ & $-1$\\
3 & $1$ & $1$ & $1$ & $1$\\
5 & $2+\sqrt3$ & $-138.2820$ & $11$ & $-7898.0758$\\
7 & $3+2\sqrt2$ & $-950.1758$ & $29$ & $-194822.0302$\\
11 & $5+2\sqrt6$ & $-8622.1019$ & $89$ & $-8458382.0103$
\end{tabular}
\end{table}

\section{Concluding remarks}\label{sec:concl}

In this work, we have written the $p$-adic power-law integral identity as a star-triangle relation and constructed from it a unitary Yang-Baxter operator~\eqref{eq:R}. This operator is the main result of the paper, and its properties are collected in Theorem~\ref{thm:main}. The calculation follows the operator method of~\cite{Gahramanov:2015cva,Gahramanov:2022jxz}, where the local gamma function provides the normalization. The kernel of the operator is a product of four power-law weights. For imaginary spectral and spin parameters, it intertwines the Hilbert tensor products of principal series representations of $\GL$ and satisfies the Yang-Baxter equation. For equal spins it reduces to the identity at $\delta=0$.

The same operator form of the star-triangle relation gives a Bailey lemma and a Bailey chain. Starting from a seed which is computed by the star-triangle relation, we have obtained the transformation formula~\eqref{eq:fourpointtrans} for four-point integrals. Since integrals of products of powers over $\Qp$ reduce to finite sums of geometric series, the identities obtained in this way are elementary.

The continuous weights themselves do not satisfy Wu's relations. Therefore we applied Wu's construction to a finite model obtained from the same weights. Averaging over the first residue level of $\mathbb P^1(\Qp)$ gives Potts weights with $p+1$ spin values. In this model the convolution relation pairs the exponents $a$ and $2-a$, and the pointwise relation fixes $t_p+t_p^{-1}=p-1$. After choosing the crossing normalization, the state sum~\eqref{eq:padic-jones-sum} gives $V_K(t_p)$. Note that the required exponents~\eqref{eq:exponents} contain the unramified quadratic character and use the analytic continuation of the pair averages. The Potts form of the averaged weights follows from their invariance under $PGL_2(\Zp)$, see Lemma~\ref{lem:potts}. Therefore the state sum is the known Potts evaluation of the Jones polynomial at $q=p+1$, and the finite star-triangle relation is not derived from the continuous one. Although the relation between the Potts model and the Jones polynomial is well known, it is interesting that the $p$-adic weights lead to this particular evaluation in such a direct way. There are several directions that we wish to pursue in the future.

For identical imaginary spin parameters and imaginary $\delta$ the operator $\Rc(\delta)$ commutes with the action of $\GL$ on $\mathcal H_{\frac12+g}\widehat\otimes\mathcal H_{\frac12+g}$. It would be interesting to find its spectral decomposition, i.e.\ the eigenvalues of $\Rc(\delta)$ on the irreducible components of the tensor product, as was done for $SL(2,\mathbb C)$ in~\cite{Derkachov:2006fw}. This would also clarify the analytic properties of the operator in the spectral parameter.

Kels and Yamazaki obtained finite spin models by taking a root-of-unity limit of the lens elliptic weights~\cite{Kels:2017vbc}. It would be interesting to compare that procedure with the averaging used here.

One can also average over finer residue levels, where the weights distinguish several distances. At level $n$ the spins take values in $\mathbb P^1(\mathbb Z/p^n\mathbb Z)$. For $n\ge2$ the group $PGL_2(\mathbb Z/p^n\mathbb Z)$ is not doubly transitive on this set, and its orbits on pairs of points are labeled by the $n+1$ values of the chordal distance. Hence the averaged weights take $n+1$ values, and the corresponding model is no longer the Potts model. It would be interesting to understand which link invariants arise in this way, and whether they are related to the $p$-adic knot polynomials proposed by Morozov~\cite{Morozov:2015nra}. The chiral Potts construction of Wu, Pant and King~\cite{Wu:1994chiral,Wu:1995chiral} gives another possible direction for finding link invariants from such finite models.

For the continuous operator, the main open question is the product of the kernels obtained at the crossing point (Appendix~\ref{app:crossing}). It seems that these kernels generate a Temperley-Lieb algebra with zero loop parameter, but we do not yet know how to construct a common regularization and a reduced closure. Establishing them would give a direct connection between the Yang-Baxter operator~\eqref{eq:R} and the Jones evaluation at $t=-1$.

Garoufalidis and Kashaev constructed quantum dilogarithms satisfying a pentagon identity over local fields~\cite{Garoufalidis:2023dhd}. It would be interesting to understand whether the power-law star-triangle relation of the present model can be connected to that identity or to a pentagon relation of the type studied in~\cite{Bozkurt:2020gyy,Gahramanov:2022jxz}. It would also be interesting to look for $p$-adic analogues of the star-star relation~\cite{Catak:2021coz,Catak:2025hoz}.

\appendix
\renewcommand{\thesection}{\Alph{section}}
\titleformat{\section}{\centering\normalfont\bfseries}{APPENDIX \thesection.}{.5em}{\MakeUppercase}

\section{Operator relations in the unitary case}\label{app:unitary}

In this appendix, we prove Lemma~\ref{lem:opstr} and the statements of Section~\ref{sec:yb} concerning the operators $M(a)$. As in the real principal-series analysis of Neretin~\cite{Neretin:2024}, we first establish the identity on test functions and then pass to unitary operators by density. Let $\mathcal D(\Qp)$ be the space of locally constant functions on $\Qp$ with compact support. We use the Fourier transform and the bilinear pairing
\begin{equation}\label{eq:pairing}
 \hat f(k)=\int_{\Qp}f(x)\,\psi(kx)\,dx,\qquad
 (f,g)=\int_{\Qp}f(x)\,g(x)\,dx=\int_{\Qp}\hat f(-k)\,\hat g(k)\,dk .
\end{equation}
The Fourier transform maps $\mathcal D(\Qp)$ onto itself~\cite{Vladimirov:1994wi}. The operator $M(a)$ is defined by $\widehat{M(a)f}(k)=|k|_p^{-a}\hat f(k)$, so that
\begin{equation}\label{eq:Msym}
 (f,M(a)g)=\int_{\Qp}\hat f(-k)\,|k|_p^{-a}\,\hat g(k)\,dk=(M(a)f,g).
\end{equation}
For $f,g\in\mathcal D(\Qp)$ this integral converges for $\operatorname{Re}a<1$. If $0<\operatorname{Re}a<1$ and $g\in\mathcal D(\Qp)$, then by~\eqref{eq:fourier} the function $M(a)g$ is given by the convergent integral~\eqref{eq:MD}.

\paragraph{Functions with vanishing integral} Let $f\in\mathcal D(\Qp)$ and $\int_{\Qp}f\,dx=0$. Then $\hat f\in\mathcal D(\Qp)$ and $\hat f(0)=0$. Since $\hat f$ is locally constant, it vanishes in a neighborhood of $k=0$. Therefore the function $|k|_p^{-a}\hat f(k)$ belongs to $\mathcal D(\Qp)$ and vanishes near $k=0$ for all $a\in\mathbb C$. Hence $M(a)f\in\mathcal D(\Qp)$, the integral of $M(a)f$ vanishes, and $M(a)M(b)f=M(a+b)f$ for all $a,b\in\mathbb C$. Moreover, $\hat f$ is a finite linear combination of characteristic functions of discs which do not contain the point $k=0$, and $|k|_p$ is constant on each of them. Thus
\begin{equation}\label{eq:Mfinite}
 M(a)f=\sum_j p^{an_j}\,e_j,
\end{equation}
where the sum is finite, $n_j\in\mathbb Z$, and the functions $e_j\in\mathcal D(\Qp)$ do not depend on $a$.

The space $\Phi_y$ of Lemma~\ref{lem:opstr} is dense in $L^2(\Qp)$. Indeed, the functions from $\mathcal D(\Qp)$ which vanish near $y$ are dense. Let $f$ be such a function, and let $S_n=\{x:|x-y|_p=p^n\}$ be the sphere of measure $p^n(1-p^{-1})$. Then the function
\begin{equation}
 f_n=f-\frac{\int_{\Qp}f\,dx}{p^n(1-p^{-1})}\,\mathbf 1_{S_n}
\end{equation}
belongs to $\Phi_y$, and $\|f-f_n\|^2=\bigl|\int_{\Qp}f\,dx\bigr|^2p^{-n}(1-p^{-1})^{-1}\to0$ as $n\to\infty$.

\paragraph{Proof of (i)} Let $f,g\in\Phi_y$, and let us denote by $L(s,t)$ and $R(s,t)$ the left-hand side and the right-hand side of~\eqref{eq:opstr-matrix}, respectively. By~\eqref{eq:Mfinite}, the product $(M(s)f)(x)\,(M(t)g)(x)$ is a finite linear combination of characteristic functions of discs, with coefficients which are entire functions of $s$ and $t$. If a disc $B$ does not contain $y$, then $|x-y|_p$ is constant on $B$ and $\int_B|x-y|_p^{-s-t}dx$ is an entire function of $s+t$. For the discs centered at $y$ one has
\begin{equation}\label{eq:disc-int}
 \int_{|x-y|_p\le p^{-m}}|x-y|_p^{-c}\,dx=\frac{(1-p^{-1})\,p^{-m(1-c)}}{1-p^{c-1}},\qquad \operatorname{Re}c<1.
\end{equation}
Hence $L(s,t)$ is analytic for $\operatorname{Re}(s+t)<1$. Further, since $f$ and $g$ vanish near $y$, the functions
\begin{equation}
 f_t(x)=|x-y|_p^{-t}f(x),\qquad g_s(x)=|x-y|_p^{-s}g(x)
\end{equation}
belong to $\mathcal D(\Qp)$, and they are finite linear combinations of characteristic functions of discs with coefficients entire in $t$ and $s$, respectively. By~\eqref{eq:Msym},
\begin{equation}\label{eq:Rst}
 R(s,t)=(f_t,M(s+t)g_s)=\int_{\Qp}\hat f_t(-k)\,|k|_p^{-s-t}\,\hat g_s(k)\,dk .
\end{equation}
The product $\hat f_t(-k)\hat g_s(k)$ is again a finite linear combination of characteristic functions of discs with entire coefficients, and the same argument shows that $R(s,t)$ is analytic for $\operatorname{Re}(s+t)<1$.

Let now $\operatorname{Re}s>0$, $\operatorname{Re}t>0$ and $\operatorname{Re}(s+t)<1$. Then $M(s)f$ and $M(t)g$ are given by the integral~\eqref{eq:MD}, and
\begin{equation}\label{eq:Ltriple}
 L(s,t)=\frac{1}{\Gamma_p(s)\Gamma_p(t)}\int_{\Qp^3}f(x)\,g(w)\,|x-z|_p^{s-1}\,|z-y|_p^{-s-t}\,|z-w|_p^{t-1}\,dz\,dx\,dw .
\end{equation}
This integral converges absolutely. Indeed, let $\sigma_1=\operatorname{Re}s$ and $\sigma_2=\operatorname{Re}t$. Replacing all functions in~\eqref{eq:Ltriple} by their absolute values and integrating over $z$ by means of~\eqref{eq:rawstr} with the real exponents $(a_1,a_2,a_3)=(\sigma_1,1-\sigma_1-\sigma_2,\sigma_2)$, one obtains
\begin{equation}
 \Gamma_p(\sigma_1)\Gamma_p(\sigma_2)\Gamma_p(1-\sigma_1-\sigma_2)\int_{\Qp^2}|f(x)|\,|g(w)|\,|x-y|_p^{-\sigma_2}\,|w-y|_p^{-\sigma_1}\,|x-w|_p^{\sigma_1+\sigma_2-1}\,dx\,dw<\infty,
\end{equation}
since $f$ and $g$ have compact supports which do not contain $y$, and $\sigma_1+\sigma_2>0$. Therefore one can change the order of integration in~\eqref{eq:Ltriple}. Integrating first over $z$ by means of~\eqref{eq:rawstr} and using the reflection formula~\eqref{eq:reflection}, we obtain
\begin{equation}
 L(s,t)=\int_{\Qp^2}f_t(x)\,\frac{|x-w|_p^{s+t-1}}{\Gamma_p(s+t)}\,g_s(w)\,dx\,dw=(f_t,M(s+t)g_s)=R(s,t).
\end{equation}
Here we used once more that for $0<\operatorname{Re}(s+t)<1$ the operator $M(s+t)$ acts on $g_s\in\mathcal D(\Qp)$ by the integral~\eqref{eq:MD}. Thus $L$ and $R$ are analytic functions on the connected domain $\operatorname{Re}(s+t)<1$ in $\mathbb C^2$, and they coincide on its open subset. Hence they coincide on the whole domain. Let us also note that, by~\eqref{eq:Mfinite},~\eqref{eq:disc-int} and~\eqref{eq:Rst}, both sides of~\eqref{eq:opstr-matrix} are meromorphic functions of $(s,t)\in\mathbb C^2$ with possible poles at $s+t\equiv1$ modulo $\tau$, and the relation holds for all $s,t$ in this sense.

\paragraph{Proof of (ii)} Let $s,t\in i\mathbb R$. Then all operators in~\eqref{eq:opstr} are unitary on $L^2(\Qp)$, and~\eqref{eq:Msym} holds for all $f,g\in L^2(\Qp)$. Therefore for $f,g\in\Phi_y$
\begin{equation}
 L(s,t)=\bigl(f,M(s)D(s+t)M(t)g\bigr),\qquad R(s,t)=\bigl(f,D(t)M(s+t)D(s)g\bigr).
\end{equation}
By part (i), these matrix elements coincide. Since $\Phi_y$ is dense in $L^2(\Qp)$ and is invariant under complex conjugation, the two bounded operators coincide.

\section{Proof of the intertwining relations}\label{app:proof}

In this appendix, we prove Lemma~\ref{lem:intertwiners}. We will need two elementary identities for the Möbius action $h\cdot x=(\mathsf ax+\mathsf b)/(\mathsf cx+\mathsf d)$,
\begin{equation}\label{eq:mobius}
 |h\cdot x-h\cdot y|_p=\frac{|\det h|_p\,|x-y|_p}{|\mathsf cx+\mathsf d|_p\,|\mathsf cy+\mathsf d|_p},\qquad
 d(h\cdot x)=\frac{|\det h|_p}{|\mathsf cx+\mathsf d|_p^{2}}\,dx.
\end{equation}
The first identity follows from $(\mathsf ax+\mathsf b)(\mathsf cy+\mathsf d)-(\mathsf ay+\mathsf b)(\mathsf cx+\mathsf d)=\det h\,(x-y)$, and the second one is its limit $y\to x$.

The relation~\eqref{eq:Dint} follows directly from the first identity in~\eqref{eq:mobius}. Indeed, the factor $|h\cdot x_1-h\cdot x_2|_p^{-a}$ produces $|\det h|_p^{-a}|\mathsf cx_1+\mathsf d|_p^{a}|\mathsf cx_2+\mathsf d|_p^{a}$, which shifts both parameters by $a/2$. In order to prove the relation~\eqref{eq:Mint}, let us change the integration variable in $M(a)T_\rho(h)f$ to $w=h\cdot z$, and put $X=h\cdot x$ and $j(w)=-\mathsf cw+\mathsf a$. Applying~\eqref{eq:mobius} to the inverse transformation $z=(\mathsf dw-\mathsf b)/j(w)$, one finds
\begin{equation}
 |\mathsf cz+\mathsf d|_p=\frac{|\det h|_p}{|j(w)|_p},\qquad
 dz=\frac{|\det h|_p}{|j(w)|_p^{2}}\,dw,\qquad
 |x-z|_p=\frac{|\mathsf cx+\mathsf d|_p\,|X-w|_p}{|j(w)|_p}.
\end{equation}
Therefore
\begin{equation}
 \bigl(M(a)T_\rho(h)f\bigr)(x)=|\det h|_p^{1-\rho}\,|\mathsf cx+\mathsf d|_p^{a-1}
 \int_{\Qp}\frac{|X-w|_p^{a-1}}{\Gamma_p(a)}\,|j(w)|_p^{2\rho-1-a}\,f(w)\,dw.
\end{equation}
For $2\rho=1+a$ the factor $|j(w)|_p$ drops out, and the right-hand side becomes $|\det h|_p^{(1-a)/2}|\mathsf cx+\mathsf d|_p^{-(1-a)}(M(a)f)(h\cdot x)$, which is~\eqref{eq:Mint}.

\paragraph{Unitary case} The relation~\eqref{eq:Dint} holds pointwise almost everywhere for all $a$. For $a\in i\mathbb R$ the integral~\eqref{eq:MD} does not converge, and the calculation above should be supplemented by the following argument. We use the notation of Appendix~\ref{app:unitary}. The operators $M(a)$ and $T_\rho(h)$ with $\operatorname{Re}\rho=\tfrac12$ are unitary on $L^2(\Qp)$. Since $T_\rho(h_1)T_\rho(h_2)=T_\rho(h_2h_1)$, it is sufficient to check~\eqref{eq:Mint} for the generators of $\GL$, i.e.\ for the upper triangular matrices and for the matrix $w_0=\left(\begin{smallmatrix}0&1\\1&0\end{smallmatrix}\right)$. For an upper triangular matrix one has $(T_\rho(h)f)(x)=|\lambda|_p^{\rho}f(\lambda x+\mu)$ with $\lambda=\mathsf a/\mathsf d$ and $\mu=\mathsf b/\mathsf d$. The multiplication by $|k|_p^{-a}$ in momentum space commutes with translations, and for the dilation $f^\lambda(x)=f(\lambda x)$ one finds $M(a)f^\lambda=|\lambda|_p^{-a}(M(a)f)^\lambda$, which gives~\eqref{eq:Mint}. For the matrix $w_0$ one has $(T_\rho(w_0)f)(x)=|x|_p^{-2\rho}f(1/x)$. Let $f,g\in\mathcal D(\Qp)$ vanish in a neighborhood of zero. Then $T_\rho(w_0)f$ and $T_\rho(w_0)g$ belong to $\mathcal D(\Qp)$ and are entire in $\rho$. Hence, by~\eqref{eq:Msym}, the functions
\begin{equation}
 P(a)=\bigl(f,M(a)T_{\frac{1+a}2}(w_0)g\bigr),\qquad Q(a)=\bigl(T_{\frac{1+a}2}(w_0)f,M(a)g\bigr)
\end{equation}
are analytic for $\operatorname{Re}a<1$. For $0<\operatorname{Re}a<1$ all integrals converge absolutely, and the calculation above gives $P(a)=\bigl(f,T_{\frac{1-a}2}(w_0)M(a)g\bigr)$. The change of variable $x\to1/x$ in the latter integral gives $Q(a)$. Therefore $P(a)=Q(a)$ for $\operatorname{Re}a<1$. For $a\in i\mathbb R$ the same change of variable gives $Q(a)=\bigl(f,T_{\frac{1-a}2}(w_0)M(a)g\bigr)$. Since the functions from $\mathcal D(\Qp)$ which vanish near zero are dense in $L^2(\Qp)$, we arrive at~\eqref{eq:Mint}.

\section{Wu's edge-interaction construction}\label{app:wu}

In this appendix, we briefly review the edge-interaction construction of Wu~\cite{Wu:1992zzb}. Let us color the faces of a link diagram as a checkerboard and place a spin on each shaded face. A crossing carries the weight $W_+$ or $W_-$ according to the shaded-face convention of Wu~\cite{Wu:1992zzb}.\footnote{These labels should not be identified with the oriented crossing signs which define the writhe.} Equivalently, the shaded faces are the vertices of a graph and each crossing gives an edge joining the two adjacent shaded faces.

Let $q$ be the number of spin states. We introduce a normalization constant $\mathcal D$ and define the partition function by
\begin{equation}\label{eq:Z}
 \mathcal D^2=q,\qquad Z=\mathcal D^{-N}\sum_{\text{spins}}\;\prod_{\text{crossings}}W_\pm(\sigma,\sigma'),
\end{equation}
where $N$ is the number of shaded faces and the sum runs over all assignments of the $q$ spin values to the vertices. For symmetric weights $W_\pm(\sigma,\sigma')=W_\pm(\sigma',\sigma)$, the edge-interaction conditions have the following form
\begin{align}
 \mathcal D^{-1}\sum_{\rho}W_\pm(\sigma,\rho)&=\alpha^{\mp1},\qquad
 W_\pm(\sigma,\sigma)=\alpha^{\pm1},\tag{W1}\label{eq:W1}\\
 \sum_{\rho}W_-(\sigma_1,\rho)\,W_+(\rho,\sigma_2)&=q\,\delta_{\sigma_1\sigma_2},\tag{W2}\label{eq:W2}\\
 W_+(\sigma_1,\sigma_2)\,W_-(\sigma_1,\sigma_2)&=1,\tag{W3}\label{eq:W3}\\
 \mathcal D^{-1}\sum_{\rho}W_\pm(\sigma_1,\rho)\,W_\mp(\sigma_2,\rho)\,W_\pm(\sigma_3,\rho)
 &=W_\mp(\sigma_1,\sigma_2)\,W_\mp(\sigma_2,\sigma_3)\,W_\pm(\sigma_3,\sigma_1).\tag{W4}\label{eq:W4}
\end{align}
Let us briefly explain the meaning of these relations. The two inversion relations~\eqref{eq:W2} and~\eqref{eq:W3} remove a pair of crossings in the two checkerboard versions of the second Reidemeister move. The relation~\eqref{eq:W4} is the star-triangle relation for the third move, where the factor $\mathcal D^{-1}$ accounts for the shaded face summed over on the star side. Finally,~\eqref{eq:W1} controls the two ways of adding a curl. As a result, the partition function is invariant under the second and third moves, while the first move contributes a scalar factor. With our checkerboard labels a positive oriented curl multiplies $Z$ by $\alpha$, giving the compensating factor $\alpha^{-w(K)}$.

The solution which we need in Section~\ref{sec:jones} is the Potts model, whose weights depend only on whether the two spins are equal~\cite{Wu:1992potts}. With a consistent choice of the fourth root of $t$, the weights are
\begin{equation}\label{eq:potts}
 W_\pm(\sigma,\sigma')=t^{\pm1/4}\left[1-(1+t^{\mp1})\,\delta_{\sigma\sigma'}\right],
\end{equation}
with
\begin{equation}\label{eq:branches}
 q=t+2+t^{-1},\qquad \mathcal D=-(t^{1/2}+t^{-1/2}),\qquad
 \alpha=-t^{-3/4}.
\end{equation}
One can easily check the relations~\eqref{eq:W1} to~\eqref{eq:W4} by direct substitution. For example, for three distinct spins the plus equation in~\eqref{eq:W4} reduces to
\begin{equation}
 \mathcal D^{-1}t^{1/4}\bigl[q-2(1+t^{-1})-(1+t)\bigr]=t^{-1/4}.
\end{equation}
Note that the sign of $\mathcal D$ in~\eqref{eq:branches} is part of the normalization of the weights.

The writhe correction cancels the curl factor, while the division by $\mathcal D$ normalizes the unknot. For the Potts weights~\eqref{eq:potts} the resulting invariant is the Jones polynomial~\cite{Jones:1989ed,Wu:1992zzb,Wu:1992potts},
\begin{equation}\label{eq:jones}
 V_K(t)=\mathcal D^{-1}\alpha^{-w(K)}Z.
\end{equation}
The unknot gives $Z=\mathcal D$, hence $V_{\rm unknot}=1$. We use the graph formula for connected projections and extend to split unions by $V_{L\sqcup L'}=\mathcal D V_L V_{L'}$, where $L'$ denotes a second link. Thus the unlink of two components gives $V=\mathcal D=-(t^{1/2}+t^{-1/2})$, as required by the Jones skein relation
\begin{equation}\label{eq:skein}
 t^{-1}V_{L_+}-t\,V_{L_-}=(t^{1/2}-t^{-1/2})\,V_{L_0}.
\end{equation}

\section{Wu's relations for power-law weights}\label{app:continuous}

Let us check whether the continuous power-law weights satisfy the relations of Appendix~\ref{app:wu}, with the sums over the spins replaced by integrals over $\Qp$. We take
\begin{equation}\label{eq:cont-weights}
 W_\pm(x,y)=c_\pm|x-y|_p^{s_\pm},\qquad c_\pm\ne0.
\end{equation}

\paragraph{Inversion relations} The pointwise inversion relation~\eqref{eq:W3} gives $s_-\equiv-s_+\pmod{\tau}$ and $c_+c_-=1$. To check the second one, we use the Fourier transform
\begin{equation}
 \widehat{|x|_p^s}(k)=\Gamma_p(s+1)\,|k|_p^{-s-1},
\end{equation}
which is~\eqref{eq:fourier} with $a=s+1$. For generic exponents, the Fourier transform of the convolution of $W_+$ and $W_-$ is then proportional to $|k|_p^{-2}$. Thus the convolution does not give the delta function in~\eqref{eq:W2}, and the two inversion relations cannot hold simultaneously. For example, the pointwise inverse of $|x|_p^{-a}$ is $|x|_p^a$, while its convolution inverse is proportional to $|x|_p^{a-2}$. Note that this does not contradict the inversion relations~\eqref{eq:inv2} of Section~\ref{sec:str}. Indeed, there the pointwise and the convolution relations involve two different weights, $W$ and $\overline W$, related by the crossing $a\to1-a$, while in Wu's construction both relations must hold for the same pair of weights $W_\pm$.

\paragraph{Star-triangle relation} For the weights~\eqref{eq:cont-weights} the plus equation in~\eqref{eq:W4} reads
\begin{equation}\label{eq:cont-str}
 \mathcal D^{-1}\int_{\Qp}W_+(x_1,x_0)\,W_-(x_2,x_0)\,W_+(x_3,x_0)\,dx_0=W_-(x_1,x_2)\,W_-(x_2,x_3)\,W_+(x_3,x_1).
\end{equation}
The integral converges for $\operatorname{Re}s_\pm>-1$ and $\operatorname{Re}(2s_++s_-)<-1$. The relation~\eqref{eq:cont-str} is not a particular case of the star-triangle relation~\eqref{eq:rawstr}. Indeed, its left-hand side is the integral in~\eqref{eq:rawstr} with $a_1=a_3=s_++1$ and $a_2=s_-+1$. For $s_-=-s_+$ the balancing condition gives $s_+=-2$. Then the right-hand side of~\eqref{eq:rawstr} is formally proportional to $|x_{12}|_p\,|x_{23}|_p\,|x_{13}|_p^{-3}$, whereas the right-hand side of~\eqref{eq:cont-str} is proportional to $|x_{12}|_p^{2}\,|x_{23}|_p^{2}\,|x_{13}|_p^{-2}$. Proposition~\ref{prop:nostr} below rules out~\eqref{eq:cont-str} throughout its convergence region. For its proof we use the unbalanced three-point formula of Gubser, Jepsen, Parikh and Trundy~\cite{Gubser:2017vgc} in one dimension. We derive it by summing over shells in order to keep track of the coefficients and of the resonant case used below.

\begin{lemma}\label{lem:unbalanced}
Let $\operatorname{Re}a_j>0$, $\operatorname{Re}(a_1+a_2+a_3)<2$ and $p^{a_1+a_2-1}\ne1$, and let $|x_{13}|_p=|x_{23}|_p>|x_{12}|_p$. Then
\begin{equation}\label{eq:unbalanced}
 \int_{\Qp}\prod_{j=1}^3|x_j-x_0|_p^{a_j-1}\,dx_0=C_1\,|x_{12}|_p^{a_1+a_2-1}|x_{13}|_p^{a_3-1}+C_2\,|x_{13}|_p^{a_1+a_2+a_3-2},
\end{equation}
where
\begin{equation}\label{eq:C12}
\begin{split}
 C_1&=\tfrac1p\bigl[r(1-a_1)+r(1-a_2)-r(2-a_1-a_2)-1\bigr],\\
 C_2&=\tfrac1p\bigl[r(2-a_1-a_2)+r(1-a_3)-r(3-a_1-a_2-a_3)-1\bigr],
\end{split}
\end{equation}
and $r(a)=(p-1)/(p^{1-a}-1)$ is the function~\eqref{eq:r}.
\end{lemma}

\begin{proof}
Let $\ell=|x_{12}|_p$ and $\ell'=|x_{13}|_p$. Summing over the spheres $|u|_p=p^n$ of measure $(1-p^{-1})p^n$, one finds
\begin{equation}\label{eq:ball-ints}
 \int_{|u|_p<\ell}|u|_p^{a-1}du=\frac{r(1-a)}{p}\,\ell^{a}\quad(\operatorname{Re}a>0),\qquad
 \int_{|u|_p>\ell}|u|_p^{a-1}du=-\frac{p-1+r(1-a)}{p}\,\ell^{a}\quad(\operatorname{Re}a<0),
\end{equation}
and, for $p^a\ne1$,
\begin{equation}\label{eq:shell-int}
 \int_{\ell<|u|_p<\ell'}|u|_p^{a-1}du=\frac{r(1-a)}{p}\,\ell'^{\,a}-\frac{p-1+r(1-a)}{p}\,\ell^{a}.
\end{equation}
Let us split the integral in~\eqref{eq:unbalanced} according to the position of $x_0$. If $|x_0-x_1|_p<\ell$, then $|x_0-x_2|_p=\ell$ and $|x_0-x_3|_p=\ell'$, and~\eqref{eq:ball-ints} gives $\ell^{a_1+a_2-1}\ell'^{\,a_3-1}r(1-a_1)/p$. The region $|x_0-x_2|_p<\ell$ gives the same expression with $r(1-a_2)$. The set $|x_0-x_1|_p=|x_0-x_2|_p=\ell$ has the measure $(p-2)\ell/p$ and gives $\ell^{a_1+a_2-1}\ell'^{\,a_3-1}(p-2)/p$. If $\ell<|x_0-x_1|_p<\ell'$, then $|x_0-x_2|_p=|x_0-x_1|_p$ and $|x_0-x_3|_p=\ell'$, and this region gives $\ell'^{\,a_3-1}$ times the integral~\eqref{eq:shell-int} with $a=a_1+a_2-1$. If $|x_0-x_1|_p=\ell'$, then $|x_0-x_2|_p=\ell'$ and $|x_0-x_3|_p\le\ell'$, and in the same way as above one obtains $\ell'^{\,a_1+a_2+a_3-2}\,[r(1-a_3)+p-2]/p$. Finally, if $|x_0-x_1|_p>\ell'$, then all three distances are equal, and the second formula in~\eqref{eq:ball-ints} with $a=a_1+a_2+a_3-2$ applies. Collecting all terms, we arrive at~\eqref{eq:unbalanced}.
\end{proof}

For $a_1+a_2+a_3=1$ the identity~\eqref{eq:r-identity} and $r(2)=-p$ give $C_2=0$ and $C_1=\frac1p[r(1-a_1)+r(1-a_2)+r(1-a_3)+p-2]$. A direct calculation shows that the latter expression is equal to $\Gamma_p(a_1)\Gamma_p(a_2)\Gamma_p(a_3)$, so that~\eqref{eq:unbalanced} reduces to~\eqref{eq:rawstr}. Without the balancing condition the second term in~\eqref{eq:unbalanced} does not vanish in general, and the integral is not a product of powers of the distances.

\begin{proposition}\label{prop:nostr}
The weights~\eqref{eq:cont-weights} do not satisfy the relation~\eqref{eq:cont-str} for any values of $c_\pm$, $\mathcal D$ and of the exponents $s_\pm$ for which the integral converges.
\end{proposition}

\begin{proof}
Let $|x_{13}|_p=|x_{23}|_p=p^n$ and $|x_{12}|_p=p^m$, where $m<n$ are arbitrary integers. Let us first assume that $p^{s_++s_-+1}\ne1$. By Lemma~\ref{lem:unbalanced} with $a_1=a_3=s_++1$ and $a_2=s_-+1$, the relation~\eqref{eq:cont-str} takes the form
\begin{equation}
 \mathcal D^{-1}c_+^2c_-\Bigl[C_1\,p^{m(s_++s_-+1)}p^{ns_+}+C_2\,p^{n(2s_++s_-+1)}\Bigr]=c_+c_-^2\,p^{ms_-}p^{n(s_++s_-)}.
\end{equation}
The functions $p^{m\mu+n\nu}$ of the integers $m<n$ with different pairs $(p^\mu,p^\nu)$ are linearly independent, and the right-hand side does not vanish. Therefore the function on the right-hand side must coincide with one of the two functions on the left-hand side, i.e.\ either $p^{s_-}=p^{s_++s_-+1}$ and $p^{s_++s_-}=p^{s_+}$, or $p^{s_-}=1$ and $p^{s_++s_-}=p^{2s_++s_-+1}$. In both cases
\begin{equation}\label{eq:trivial-point}
 p^{s_-}=1,\qquad p^{s_++1}=1.
\end{equation}
Then $\operatorname{Re}s_+=-1$, and the integral in~\eqref{eq:cont-str} diverges. If $p^{s_++s_-+1}=1$, then the integral~\eqref{eq:shell-int} is replaced by $(1-p^{-1})(n-m-1)$. In this case the left-hand side of~\eqref{eq:cont-str} depends on $m$ through a linear function with a nonvanishing coefficient, while the right-hand side is proportional to $p^{ms_-}$. Therefore the relation does not hold in this case either.
\end{proof}

At the exceptional parameters~\eqref{eq:trivial-point}, the normalized Riesz kernel $|x-y|_p^{s_+}/\Gamma_p(s_++1)$ specializes to $\delta(x-y)$, while $|x-y|_p^{s_-}=1$ off the diagonal. Let us write $W_+=\gamma_+\delta(x-y)$ and $W_-=\gamma_-$ with constants $\gamma_\pm\ne0$. The plus branch of~\eqref{eq:W4} then reads $\mathcal D^{-1}\gamma_+^2\gamma_-\,\delta(x_1-x_3)=\gamma_+\gamma_-^2\,\delta(x_1-x_3)$, which holds precisely when $\gamma_+=\mathcal D\gamma_-$. The minus branch would require
\[
 \mathcal D^{-1}\gamma_+\gamma_-^2=\gamma_+^2\gamma_-\,\delta(x_1-x_2)\delta(x_2-x_3).
\]
Pairing with the indicator of a product of three disjoint balls gives a nonzero value on the left and zero on the right. Thus this singular pair fails the full relation~\eqref{eq:W4}. Its pointwise product also fails~\eqref{eq:W3}.

\paragraph{Quasi-characters} One can also consider weights proportional to a quasi-character $\pi(x-y)$ and its inverse. Rescaling the three external spins in~\eqref{eq:W4} by $\lambda$ gives
\begin{equation}
 |\lambda|_p\,\pi(\lambda)=\pi(\lambda)^{-1}.
\end{equation}
Hence $\pi=\chi|\cdot|_p^{-1/2}$ with $\chi^2=1$. If the character $\chi$ is unramified, i.e.\ $\chi=1$ or $\chi=\eta$, then these weights are of the form~\eqref{eq:cont-weights}, and the analysis above applies. In any case, this condition does not resolve the second inversion relation. Note that these observations concern generic exponents, and the special values at which the kernels become distributions should be treated separately. This is why in Section~\ref{sec:finite} we turn to a finite model obtained by averaging the weights over residue discs.

\section{Crossing-point limit and the Temperley-Lieb algebra}\label{app:crossing}

In this appendix, we examine the Yang-Baxter operator~\eqref{eq:R} at the crossing point $\delta=1$. The construction of braid representations from Boltzmann weights and the use of Markov traces were studied in~\cite{Akutsu:1987qs,Deguchi:1989vd}. Wu considered an infinite-rapidity limit of the Yang-Baxter solution~\cite{Wu:1993yb}. Here we study the leading term at a finite pole and ask whether it can be used to construct a constant braid operator.

We set $g_1=g_2=g$ and assume that $\kappa_g=\Gamma_p(-2g)\Gamma_p(2g)$ is finite and nonzero. The two convolution exponents in~\eqref{eq:ABCE} are then both equal to $\delta$. Writing $\delta=1+\varepsilon$, we have, for $x\ne0$,
\begin{equation}
 \frac{|x|_p^{\varepsilon}}{\Gamma_p(1+\varepsilon)}
 =\frac{c_p}{\varepsilon}+O(1),\qquad
 c_p=-\frac{1-p^{-1}}{\ln p}.
\end{equation}
Substituting this expansion into the kernel~\eqref{eq:Rkernel}, away from its diagonals, one obtains
\begin{equation}\label{eq:residue}
 \Rc(1+\varepsilon)=\frac{c_p^2}{\varepsilon^2}\,e+O(\varepsilon^{-1}),
 \qquad e(x_1,x_2;z_1,z_2)
 =|x_{12}|_p^{-1-2g}\,|z_{12}|_p^{-1+2g}.
\end{equation}
The dependence on the incoming and outgoing variables separates, and we get a formal rank-one kernel. In the language of Section~\ref{sec:intertwining}, the function $|x_{12}|_p^{-1-2g}$ is (formally) invariant under the diagonal action with representation labels $(\tfrac12+g,\tfrac12+g)$ away from the diagonal. The factor $|z_{12}|_p^{-1+2g}\,dz_1dz_2$ transforms as the corresponding invariant density there. The expansion is pointwise on the complement of the diagonals. Identifying the full distributional Laurent coefficient would also require control of terms supported on those diagonals. In order to multiply these kernels we need a prescription for their singular factors.

The convolution identity needed for the product is
\begin{equation}
 |x|_p^{-1-2g}*|x|_p^{-1+2g}
 =\kappa_g\,\delta(x),\qquad
 \kappa_g=\Gamma_p(-2g)\Gamma_p(2g),
\end{equation}
which follows from the normalized Riesz kernels for generic $g$. We denote by $e_i$ the kernel $e$ acting on the variables $x_i$ and $x_{i+1}$. Using this identity twice, one formally obtains $e_i e_{i+1}e_i=\kappa_g^2e_i$.

The difficulty appears in $e_i^2$, which contains the divergent integral $\int\!\int |z_1-z_2|_p^{-2}\,dz_1dz_2$. In order to compare it with the compact model of Section~\ref{sec:finite}, let us consider the chordal integral
\begin{equation}\label{eq:loop}
 I(s)=\iint_{\mathbb P^1(\Qp)^2}[z_1,z_2]^s\,d\mu(z_1)\,d\mu(z_2)
 =\frac{p}{p+1}\,\frac{1-p^{-2-s}}{1-p^{-1-s}}.
\end{equation}
This is the same sum over residue discs as in~\eqref{eq:r}. Two points lie in different discs with probability $p/(p+1)$ and in the same disc with probability $1/(p+1)$, therefore $I(s)=[p+r(-s)]/(p+1)$. The integral converges for $\operatorname{Re}s>-1$, and its meromorphic continuation gives $I(-2)=0$, while $I(0)=1$.

The vanishing of $I(-2)$ suggests a Temperley-Lieb relation with zero loop parameter. It does not yet prove $e_i^2=0$, since the convolution above uses the affine measure, whereas $I(s)$ uses the compact measure. One needs a regularization that gives both products with consistent kernel and measure factors. One also needs a compatible trace to close the braid, as in the finite-model constructions of~\cite{Deguchi:1989vd}. These steps remain to be worked out, and we hope to return to them in the future.

Suppose that this can be done and that $E_i=e_i/\kappa_g$ satisfy
\begin{equation}
 E_i^2=0,\qquad E_iE_{i\pm1}E_i=E_i,\qquad [E_i,E_j]=0\quad (|i-j|>1).
\end{equation}
Then
\begin{equation}
 b_i=\zeta\,1+\zeta^{-1}E_i,\qquad -\zeta^2-\zeta^{-2}=0,
\end{equation}
gives a constant braid operator with the inverse $\zeta^{-1}1+\zeta E_i$. This is the usual Temperley-Lieb construction, where $\zeta$ plays the role of the variable $A$ of the Kauffman bracket~\cite{Kauffman:1987}. The condition on $\zeta$ gives $t=\zeta^{-4}=-1$. Since the loop value is zero, the closure must be reduced by cutting and normalizing one component. At this value of $t$, the Jones polynomial satisfies
\begin{equation}
 |V_K(-1)|=\det K
\end{equation}
for knots. In order to obtain this evaluation from the $p$-adic Yang-Baxter operator, the regularization and the reduced closure described above must still be constructed.

\section*{ACKNOWLEDGMENTS}

The authors thank Ya\u{g}mur B\"ulb\"ul and Do\u{g}a Sa\u{g}lam for taking part in the initial stages of the project. Ilmar Gahramanov thanks Alexei Morozov for suggesting that we compute knot invariants from the solution of the star-triangle relation. Firdevs Karaku\c{s} is grateful to Igor Volovich and Irina Aref'eva for their guidance and warm hospitality during her research visit to the Steklov Mathematical Institute, where part of this work was carried out and where she learned much of the background on $p$-adic mathematical physics used here. She is also grateful to Daniil Stepanenko for many useful discussions and for his hospitality. The authors are grateful to Igor Volovich for reading the manuscript and for his valuable comments. Some of the ideas presented here were discussed at the conference ``Fields and Strings 2025'' at the Steklov Mathematical Institute (Moscow, December 2025), and we thank all the organizers, especially Dmitri Bykov and Edvard Musaev. Parts of the computations, of the code and of the drafting of this paper were carried out with the assistance of AI systems, namely Claude (Anthropic) and ChatGPT (OpenAI). Anthropic provided access to its Claude models through the Claude Team plan for scientists program. The work of Ilmar Gahramanov is supported by the Istanbul Integrability and Stringy Topics Initiative (istringy.org).

\section*{CONFLICT OF INTEREST}

The authors of this work declare that they have no conflicts of interest.

\section*{DATA AVAILABILITY}

No data were generated or analyzed in this work. The scripts used for the numerical checks mentioned in the Introduction are available from the authors upon reasonable request.

\bibliographystyle{utphys-pnuaa}
\bibliography{padic_yb_jones_pnuaa}

\end{document}